\documentclass[aps,prd,floats,floatfix,nofootinbib,superscriptaddress]{revtex4-1}
\usepackage{graphicx,amsmath,amssymb,amsfonts,hyperref,mathdots,caption,subcaption, setspace, wrapfig,slashed}
\usepackage[usenames,dvipsnames,svgnames,table]{xcolor}

\begin{document}
\raggedbottom

\title{Two viable large scalar multiplet models with a $Z_2$ symmetry}

\author{Kevin Earl}
\email{kearl@physics.carleton.ca}
\author{Katy Hartling\footnote{Formerly Katy Hally.}}
\email{khally@physics.carleton.ca}
\author{Heather E.~Logan}
\email{logan@physics.carleton.ca}
\author{Terry Pilkington}
\email{tpilking@physics.carleton.ca}

\affiliation{Ottawa-Carleton Institute for Physics, Carleton University, Ottawa, Ontario K1S 5B6, Canada}

\date{November 14, 2013}

\begin{abstract}
We study models in which the Higgs sector is extended by a single scalar electroweak multiplet $Z$ with isospin $T=5/2$ (sextet) or $7/2$ (octet) and the same hypercharge as the Standard Model Higgs doublet, in which $Z$ is odd under a global $Z_2$ symmetry.  This $Z_2$ symmetry keeps the lightest (neutral) member of $Z$ stable and has interesting implications for phenomenology.
We determine the constraints on these models from precision electroweak measurements and Higgs boson decays to two photons.  We compute the thermal relic density of the stable member of $Z$ and show that, for masses below 1~TeV, it can make up at most 1\% of the dark matter in the universe.  We also show that current dark matter direct detection experiments do not constrain the models, but future ton-scale experiments will probe their parameter space.
\end{abstract}

\maketitle
\section{Introduction}

Extensions of the scalar sector of the Standard Model (SM) beyond the minimal single Higgs doublet are of great interest in model building and collider phenomenology and are, as yet, largely unconstrained by experiment.
Such extensions are common in models that address the hierarchy problem of the SM, such as supersymmetric models~\cite{SUSY} and little Higgs models~\cite{LH}, as well as in models for neutrino masses, dark matter, etc.  Most of these extensions contain additional SU(2)$_L$-singlet, -doublet, and/or -triplet scalar fields. However, some extensions of the SM contain scalars in larger multiplets of SU(2)$_L$.  Such larger multiplets have been used to produce a natural dark matter candidate~\cite{Cirelli:2005uq,Cirelli:2009uv,Cai:2012kt}, which is kept stable thanks to an accidental global symmetry that is sometimes present in the Higgs potential for multiplets with isospin $T \geq 2$.  
Several different models with a scalar quadruplet (isospin $T=3/2$)~\cite{Babu:2009aq,Picek:2009is,Ren:2011mh,McDonald:2013kca,Law:2013gma,Law:2013saa} or scalar quintuplet ($T = 2$)~\cite{Chen:2012vm,McDonald:2013kca,Law:2013saa,McDonald:2013hsa} have also been proposed for neutrino mass generation.  The quadruplet has also been studied in the context of strengthening the electroweak phase transition~\cite{AbdusSalam:2013eya}.  Models in which the SM SU(2)$_L$-doublet Higgs mixes with a septet ($T = 3$), aided by additional representations of SU(2)$_L$, have been studied in Ref.~\cite{Hisano:2013sn}.

In this paper we consider models that extend the SM scalar sector through the addition of a \emph{single} large multiplet.  Perturbative unitarity of scattering amplitudes involving pairs of scalars and pairs of SU(2)$_L$ gauge bosons requires that $T \leq 7/2$ (i.e., $n \leq 8$) for a complex scalar multiplet and $T \leq 4$ (i.e., $n \leq 9$) for a real scalar multiplet~\cite{Hally:2012pu}.  Of particular interest are models that preserve a global U(1) or $Z_2$ symmetry under which the large multiplet is charged.  Such symmetries sometimes arise accidentally at the renormalizable level due to the structure of the scalar potential (in particular for multiplets with $T \geq 2$, i.e., of size $n \geq 5$); in other cases they can be imposed by hand.  Spontaneous breaking of a global U(1) symmetry is phenomenologically unacceptable because it would lead to a massless Goldstone boson that couples to fermions through its mixing with the CP-odd component of the SM Higgs doublet, and thus mediates new long-range forces between SM fermions.  Spontaneous breaking of a $Z_2$ symmetry can lead to problems with domain walls, as well as being tightly constrained by measurements of the rho parameter $\rho \equiv M_W^2/M_Z^2 \cos^2\theta_W \simeq 1$.  We will thus assume that the parameters of the scalar potential are chosen such that the U(1) or $Z_2$ symmetry is not spontaneously broken.  The lightest member of the single large multiplet is thus forced to be stable and becomes a dark matter candidate.  We will consider only models in which the lightest member of the large multiplet is electrically neutral; models in which the lightest member of the large multiplet is electrically charged are excluded or strongly constrained by the absence of electrically-charged relics.\footnote{Metastable multi-charged states are constrained by direct collider searches to be heavier than about 400--500~GeV, depending on their charge~\cite{ATLAS:charged}.}

The models that meet these criteria can be grouped into three classes based on the hypercharge $Y$ of the large multiplet, as follows:
\begin{enumerate}
\item[$(i)$] Models with a real, $Y=0$ multiplet, with $n = 5$, 7, or 9, corresponding to isospin $T = 2$, 3, or 4 (a real multiplet must have integer isospin).  For $n = 7$, the scalar potential preserves an accidental $Z_2$ symmetry under which the large multiplet is odd; for $n = 5$ or 9, such a $Z_2$ symmetry is not automatic but can be imposed by hand~\cite{Kumericki:2012bf}.  These models are viable and have been considered in Refs.~\cite{Cirelli:2005uq,Cirelli:2009uv} as possible candidates for ``next-to-minimal'' dark matter.  

\item[$(ii)$] Models with a complex multiplet with $n = 5$, 6, 7, or 8 (isospin $T = 2$, 5/2, 3, or 7/2), with $Y = 2T$,\footnote{We use the hypercharge normalization convention in which $Q = T^3 + Y/2$.} chosen so that the lightest member of the large multiplet can be made neutral.  The scalar potential preserves an accidental global U(1) symmetry under which the large multiplet is charged.  The masses of the states in the large multiplet are split by an operator of the form $(\Phi^{\dagger} \tau^a \Phi) (X^{\dagger} T^a X)$, where $\Phi$ is the SM Higgs doublet, $X$ is the large multiplet, and $\tau^a$ and $T^a$ are the appropriate SU(2)$_L$ generators.  We studied these models in Ref.~\cite{Earl:2013jsa} and showed that all except the $n=5$ model are excluded by dark matter direct detection experiments assuming a standard thermal history of the universe.  The model with $n=5$ avoids exclusion from this constraint because its lightest member can decay via a dimension-5 Planck-suppressed operator with a lifetime short compared to the age of the universe.

\item[$(iii)$] Models with a complex multiplet with $n = 6$ or 8 (isospin $T = 5/2$ or 7/2), with $Y = 1$.  The most general gauge-invariant and $Z_2$-invariant scalar potential for these models does not preserve any global symmetries that stabilize the large multiplet~\cite{Kumericki:2012bf}.  To obtain a dark matter candidate, we impose a $Z_2$ symmetry under which the large multiplet is odd.  The would-be global U(1) symmetry is broken by an operator of the form $(\widetilde \Phi^{\dagger} \tau^a \Phi)(Z^{\dagger} T^a \widetilde Z)$, where $\widetilde \Phi$, $\widetilde Z$ denote the conjugate multiplets.  Such an operator can appear only for this hypercharge choice and only when $n$ is even.  We study these models in the current paper.
\end{enumerate}

This paper is organized as follows.  In Sec.~\ref{sec:models} we set the notation and derive the mass eigenstates for the two $Z_2$-preserving models that we consider.  In Sec.~\ref{sec:constraints} we obtain the indirect constraints on the model parameters from perturbative unitarity and from the electroweak precision measurements via the oblique parameters $S$, $T$ and $U$.  As a by-product we present general formulas for the contributions to $S$, $T$ and $U$ from new scalar particles whose mass eigenstates are mixtures of states of definite isospin.
We also determine the contribution of the electrically-charged members of the large scalar multiplets to the loop-induced Higgs decays $h \to \gamma\gamma$ and $h \to Z \gamma$, and use the current LHC measurement of $h \to \gamma\gamma$ to further constrain the model parameters.  We also comment on the conditions to avoid alternate minima in the scalar potential in which the $Z_2$ symmetry would be spontaneously broken.  In Sec.~\ref{sec:relic} we calculate the relic density of the lightest (neutral) member of the large multiplet from thermal freeze-out.  We use this result in Sec.~\ref{sec:dirdet} to predict the dark matter direct-detection cross section and compare to the reach of current and future experiments.  We conclude in Sec.~\ref{sec:conclusions}.  Details of the spectrum calculations and Feynman rules are collected in the Appendices.  We leave an evaluation of direct LEP, Tevatron, and LHC constraints and future LHC search prospects to future work.

\section{The models}
\label{sec:models}

The models we consider here extend the SM through the addition of a single complex scalar multiplet, $Z$, with hypercharge $Y = 1$. In these models, the most general gauge-invariant scalar potential is given by
\begin{eqnarray}
	V(\Phi,Z) &=& m^2 \Phi^\dag \Phi + M^2 Z^\dag Z + \lambda_1 \big(\Phi^\dag \Phi\big)^2 
	+ \lambda_2\Phi^\dag \Phi\, Z^\dag Z + \lambda_3 \Phi^\dag \tau^a\Phi\, Z^\dag T^a Z
	\nonumber \\
	& & + \left[\lambda_4\,\widetilde{\Phi}^\dag \tau^a \Phi\;Z^\dag T^a\widetilde{Z} 
	+ \mbox{h.c.}\right] + \mathcal{O}(Z^4),
	\label{eq:conjugatepotential}
\end{eqnarray}
where $\widetilde{x} =  Cx^*$ are the conjugate multiplets.  The conjugation matrix, $C$, is an antisymmetric $n \times n$ matrix equal to $i \sigma^2$ for the SU(2)$_L$ doublet and whose form is given in Eq.~(\ref{eq:Cmatrix}) for the $n = 6$ and $n = 8$ representations. Here $\tau^a$ and $T^a$ are the generators of SU(2)$_L$ in the doublet and $n$-plet representations, respectively.  We will not need the explicit form of the $Z$ quartic couplings in what follows.

The term proportional to $\lambda_4$ is not present in the models with $Y = 2T$ considered in Ref.~\cite{Earl:2013jsa}. This term couples two SM Higgs fields $\Phi$ (not $\Phi^*$) to two $Z^*$ fields, with each pair arranged in an isospin-triplet configuration with total hypercharge $\pm 2$.  This term breaks the would-be global U(1) symmetry down to $Z_2$ and splits the masses of the real and imaginary components of the neutral member of $Z$, $\zeta^0 \equiv (\zeta^{0,r} + i \zeta^{0,i})/\sqrt{2}$.  Any complex phase of $\lambda_4$ can be absorbed into a phase rotation of $Z$ without loss of generality.  We therefore choose $\lambda_4$ to be real.

The term $Z^\dag T^a\widetilde{Z}$ is nonzero only for $T$ a half-odd integer (i.e., for even $n$).  Together with the perturbative unitarity constraint, $T \leq 7/2$ ($n \leq 8$) for complex scalar multiplets~\cite{Hally:2012pu}, this limits the models of interest to the cases $T = 5/2$ ($n = 6$) and $T = 7/2$ ($n = 8$).\footnote{The case $T = 3/2$ ($n=4$) was studied in Ref.~\cite{AbdusSalam:2013eya}.}
For these cases, the large multiplet is given in the electroweak basis by
\begin{eqnarray}
	Z_{(n=6)} &=& \left(\zeta^{+3},\,\zeta^{+2},\,\zeta^{+1},\,\zeta^{0},\,
	\zeta^{-1},\,\zeta^{-2}\right)^T, \nonumber \\
	Z_{(n=8)} &=& \left(\zeta^{+4},\,\zeta^{+3},\,\zeta^{+2},\,\zeta^{+1},\,\zeta^{0},\,
	\zeta^{-1},\,\zeta^{-2},\,\zeta^{-3}\right)^T\;.
	\label{eq:states}
\end{eqnarray}
We will denote the conjugate of the charged state $\zeta^Q$ by $\zeta^{Q*}$, which is not the same as $\zeta^{-Q}$.

We require that the lightest member of the multiplet be electrically neutral. The presence of the $\lambda_4$ term induces a mass splitting between the real and imaginary components of $\zeta^0 \equiv (\zeta^{0,r} + i \zeta^{0,i})/\sqrt{2}$, which will lead to one component being the lightest member of the multiplet and the other component being the heaviest.  Without loss of generality, we choose the real part of $\zeta^0$ to be the lightest member of the multiplet, corresponding to a choice of the sign of $\lambda_4$.  The requirement that $\zeta^{0,r}$ is lighter than any of the charged states further imposes the requirement $|\lambda_3| < 2|\lambda_4|$.

The $\lambda_4$ term induces mixing between the states $\zeta^Q$ and $\zeta^{-Q*}$ with the same (nonzero) electric charge.  The mass eigenstates are defined for $Q > 0$ in terms of mixing angles $\alpha_Q$ as
\begin{eqnarray}
	H_1^{Q} &=& \cos \alpha_Q \, \zeta^{Q} + \sin \alpha_Q \, \zeta^{-Q*},
	\nonumber \\
	H_2^{Q} &=& -\sin \alpha_Q \, \zeta^{Q} + \cos \alpha_Q \, \zeta^{-Q*},
	\label{eq:alphadef}
\end{eqnarray}
with $m_{H_1^Q} < m_{H_2^Q}$. Details of the derivation of the mass spectrum and mixing angles are given in Appendix~\ref{app:models}.

\subsection{$n=6$ model}

The mass of the real part of $\zeta^0$ in the $n=6$ model is given by
\begin{equation}
	m_{\zeta^{0,r}}^2 = M^2 + \frac{1}{2}v^2\left[\lambda_2 + \frac{1}{4}\lambda_3 + 3\lambda_4\right] 
	\equiv M^2 + \frac{1}{2}v^2\Lambda_6,
	\label{eq:lambda6}
\end{equation}
where $\Lambda_6$ is defined as the quantity in brackets above, and $v \simeq 246$ GeV is the SM Higgs vacuum expectation value (vev). The mass of the imaginary part of $\zeta^0$ is then
\begin{equation}
	m_{\zeta^{0,i}}^2 = m_{\zeta^{0,r}}^2 - 3v^2\lambda_4.
\end{equation}
Since we have chosen $\zeta^{0,r}$ to be the lightest member of the multiplet, we are forced to take $\lambda_4 < 0$.

The singly- and doubly-charged states have masses
\begin{eqnarray}
	m_{H_{1,2}^{+}}^2 &=& m_{\zeta^{0,r}}^2 + \frac{1}{4}v^2\left(-6\lambda_4 \mp \sqrt{\lambda_3^2 + 32\lambda_4^2}\right)
	\nonumber \\
	m_{H_{1,2}^{++}}^2 &=& m_{\zeta^{0,r}}^2 + \frac{1}{2}v^2\left(-3\lambda_4 \mp \sqrt{\lambda_3^2 + 5\lambda_4^2}\right),
\end{eqnarray}
where $m_{H_1^Q} < m_{H_2^Q}$ by convention.
The mixing angles for these mass eigenstates are given by
\begin{eqnarray}
	\tan \alpha_1 &=& \frac{4 \sqrt{2} \lambda_4}{\lambda_3 + \sqrt{\lambda_3^2 + 32 \lambda_4^2}}
	\nonumber \\
	\tan \alpha_2 &=& \frac{-\sqrt{5} \lambda_4}{\lambda_3 + \sqrt{\lambda_3^2 + 5 \lambda_4^2}}.
\end{eqnarray}

There is only one triply-charged state, the mass of which is given by
\begin{equation}
m_{\zeta^{+3}}^2 = m_{\zeta^{0,r}}^2 - \frac{3}{4}v^2\left(\lambda_3 + 2\lambda_4\right).
\end{equation}

\begin{figure}
\resizebox{0.49\textwidth}{!}{
\includegraphics{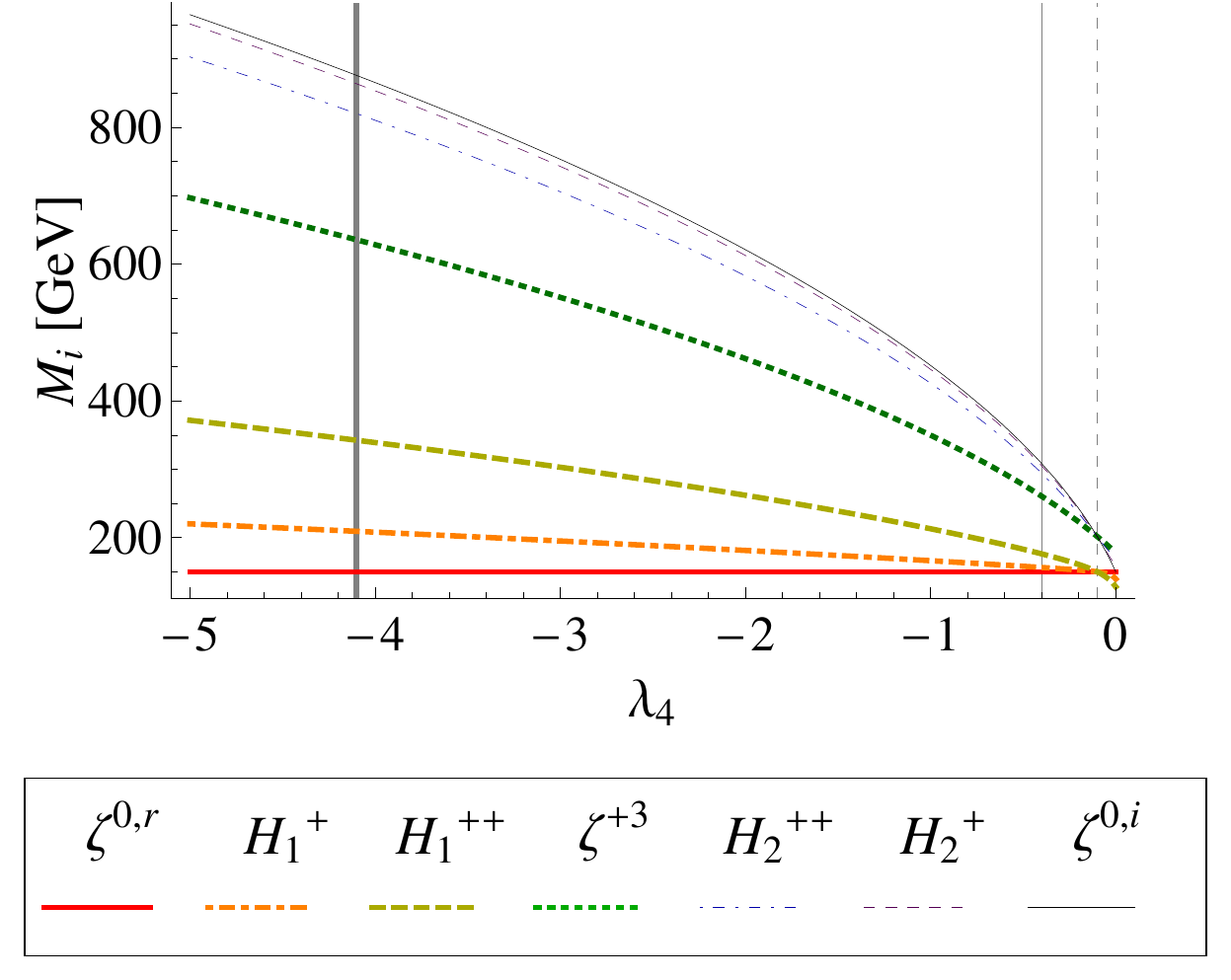}}
\resizebox{0.49\textwidth}{!}{
\includegraphics{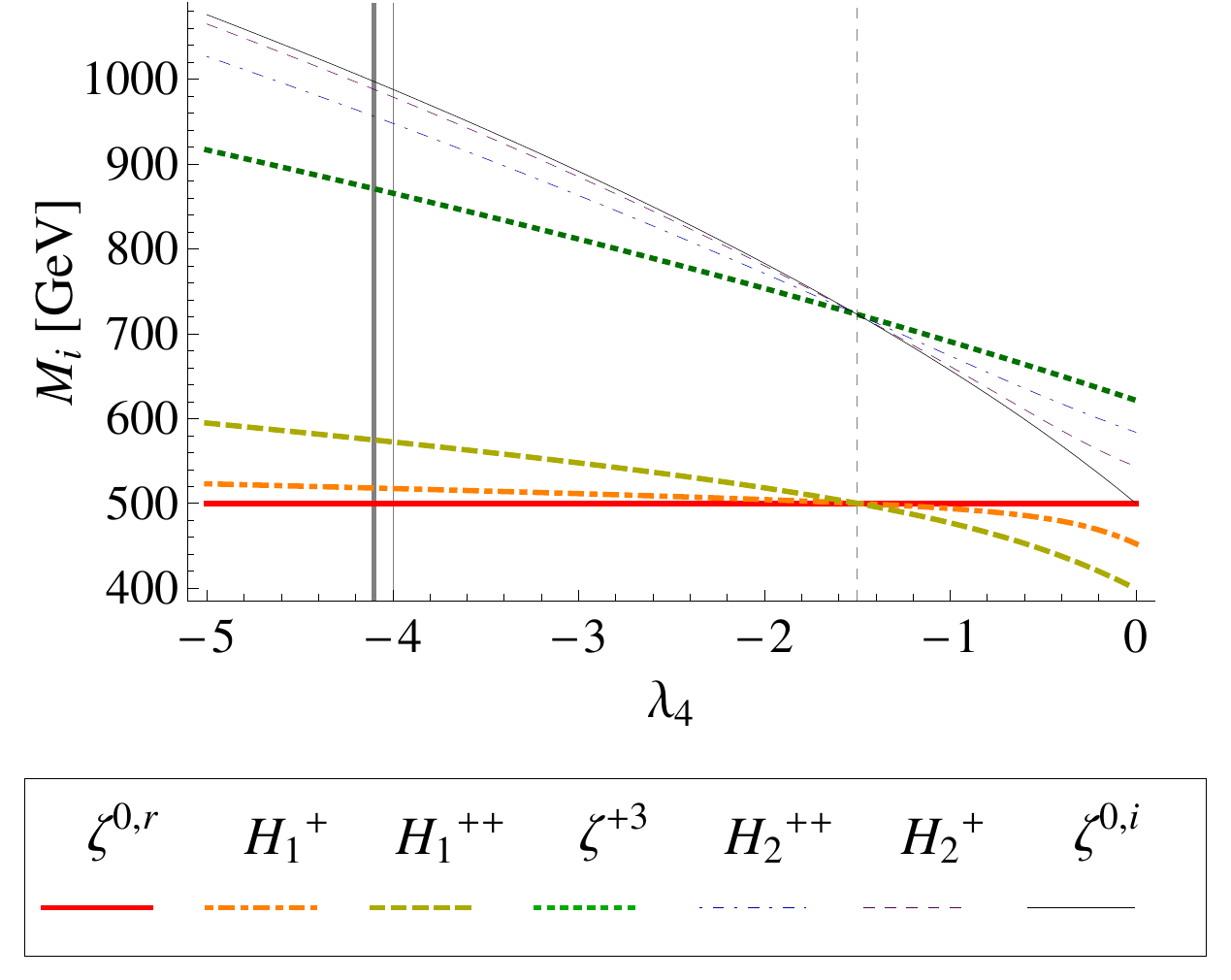}}
\caption{Sample mass spectra for the $n=6$ model as a function of $\lambda_4$.  Allowed $\lambda_4$ values lie to the left of the vertical dotted line (so that $\zeta^{0,r}$ is the lightest state) and to the right of \emph{both} vertical solid lines.  The thick vertical solid line marks the unitarity bound on $\lambda_4$ while the thin vertical solid line marks the bound from precision electroweak constraints.  Left: $m_{\zeta^{0,r}}=150$~GeV and $\lambda_3=-0.2$.   Right: $m_{\zeta^{0,r}} = 500$~GeV and $\lambda_3 = -3$.  
\label{masspec}}
\end{figure}

We note the following features of the spectrum.  When $\zeta^{0,r}$ is the lightest state, the mass eigenstates always fall in order, from lightest to heaviest, of
\begin{equation}
	\zeta^{0,r}, H_1^+, H_1^{++}, \zeta^{+3}, H_2^{++}, H_2^+, \zeta^{0,i}.
	\label{eq:6massordering}
\end{equation}
Furthermore, recall that $|\lambda_3| < 2 |\lambda_4|$ is required for $\zeta^{0,r}$ to be lighter than any of the charged states.  As we will show in Sec.~\ref{sec:STUconstraints}, constraints from the oblique parameters force $\lambda_3$ to be negative and quite close to the limit $\lambda_3 \simeq 2 \lambda_4$, unless both $\lambda_3$ and $\lambda_4$ are very small.  This leads to a clustering of the mass eigenstates in two groups: a lower-mass cluster consisting of $\zeta^{0,r}$, $H_1^+$, and $H_1^{++}$, and a higher-mass cluster consisting of $\zeta^{+3}$, $H_2^{++}$, $H_2^+$, and $\zeta^{0,i}$.  This is illustrated in Fig.~\ref{masspec}, in which we show two sample spectra as a function of $\lambda_4$, holding $m_{\zeta^{0,r}}$ and $\lambda_3$ fixed.

In the limit $\lambda_3 = 2 \lambda_4$, the mass spectrum collapses to two degenerate sets of particles: $\zeta^{0,r}$, $H_1^+$, and $H_1^{++}$ become degenerate with mass $m_{\zeta^{0,r}}$, while $\zeta^{+3}$, $H_2^{++}$, $H_2^+$, and $\zeta^{0,i}$ become degenerate with  with mass $m_{\zeta^{0,i}} = \sqrt{m_{\zeta^{0,r}}^2 - 3 v^2 \lambda_4}$.  In this limit, the composition of the mixed states in the $n=6$ model becomes
\begin{eqnarray}
	H_1^+ &=& \sqrt{\frac{1}{3}} \zeta^{+1} - \sqrt{\frac{2}{3}} \zeta^{-1*}, \qquad
	H_2^+ = \sqrt{\frac{2}{3}} \zeta^{+1} + \sqrt{\frac{1}{3}} \zeta^{-1*} \nonumber \\
	H_1^{++} &=& \sqrt{\frac{1}{6}} \zeta^{+2} + \sqrt{\frac{5}{6}} \zeta^{-2*}, \qquad
	H_2^{++} = - \sqrt{\frac{5}{6}} \zeta^{+2} + \sqrt{\frac{1}{6}} \zeta^{-2*}.
\end{eqnarray}
In particular, all transitions among the mass eigenstates mediated by $W^{\pm}$ or $Z$ emission proceed with roughly comparable coupling strength.

\subsection{$n=8$ model}

The mass of the real part of $\zeta^0$ in the $n=8$ model is given by
\begin{equation}
	m_{\zeta^{0,r}}^2 = M^2 + \frac{1}{2}v^2\left[\lambda_2 + \frac{1}{4}\lambda_3 - 4\lambda_4\right] 
	\equiv M^2 + \frac{1}{2}v^2\Lambda_8,
	\label{eq:lambda8}
\end{equation}
where $\Lambda_8$ is defined as the quantity in brackets above. The mass of the imaginary part of $\zeta^0$ is then
\begin{equation}
m_{\zeta^{0,i}}^2 = m_{\zeta^{0,r}}^2 + 4v^2\lambda_4\;.
\end{equation}
Since we have chosen $\zeta^{0,r}$ as the lightest member, we are forced to take $\lambda_4 > 0$.

The singly-, doubly-, and triply-charged states have masses
\begin{eqnarray}
	m_{H_{1,2}^{+}}^2 &=& m_{\zeta^{0,r}}^2 + \frac{1}{4}v^2\left(8\lambda_4 \mp \sqrt{\lambda_3^2 + 60\lambda_4^2}\right) \nonumber \\
	m_{H_{1,2}^{++}}^2 &=& m_{\zeta^{0,r}}^2 + \frac{1}{2}v^2\left(4\lambda_4 \mp \sqrt{\lambda_3^2 + 12\lambda_4^2}\right) \nonumber \\
	m_{H_{1,2}^{+3}}^2 &=& m_{\zeta^{0,r}}^2 + \frac{1}{4}v^2\left(8\lambda_4 \mp \sqrt{9\lambda_3^2 + 28\lambda_4^2}\right),
\end{eqnarray}
where again $m_{H_1^Q} < m_{H_2^Q}$ by convention.
The mixing angles for these mass eigenstates are given by
\begin{eqnarray}
	\tan \alpha_1 &=& \frac{-2 \sqrt{15} \lambda_4}{\lambda_3 + \sqrt{\lambda_3^2 + 60 \lambda_4^2}} \nonumber \\
	\tan \alpha_2 &=& \frac{2 \sqrt{3} \lambda_4}{\lambda_3 + \sqrt{\lambda_3^2 + 12 \lambda_4^2}} \nonumber \\
	\tan \alpha_3 &=& \frac{-2 \sqrt{7} \lambda_4}{3 \lambda_3 + \sqrt{9 \lambda_3^2 + 28 \lambda_4^2}}.
\end{eqnarray}

There is only one quadruply-charged state, the mass of which is given by
\begin{equation}
	m_{\zeta^{+4}}^2 = m_{\zeta^{0,r}}^2 - v^2\left(\lambda_3 - 2\lambda_4\right)\;.
\end{equation}

When $\zeta^{0,r}$ is the lightest state, the mass eigenstates always fall in order, from lightest to heaviest, of
\begin{equation}
	\zeta^{0,r}, H_1^+, H_1^{++}, H_1^{+3}, \zeta^{+4}, H_2^{+3}, H_2^{++}, H_2^+, \zeta^{0,i}.
\end{equation}
Constraints from the oblique parameters will again force $\lambda_3$ to be negative and quite close to the limit $\lambda_3 \simeq -2 \lambda_4$ (with $|\lambda_3| < 2 |\lambda_4|$).
In this limit, the mass spectrum again collapses to two degenerate sets of particles: $\zeta^{0,r}$, $H_1^+$, $H_1^{++}$, and $H_1^{+3}$ become degenerate with mass $m_{\zeta^{0,r}}$, while $\zeta^{+4}$, $H_2^{+3}$, $H_2^{++}$, $H_2^+$, and $\zeta^{0,i}$ become degenerate with mass $m_{\zeta^{0,i}} = \sqrt{m^2_{\zeta^{0,r}} + 4 v^2 \lambda_4}$.  In this limit, the composition of the mixed states in the $n=8$ model becomes
\begin{eqnarray}
	H_1^+ &=& \sqrt{\frac{3}{8}} \zeta^{+1} - \sqrt{\frac{5}{8}} \zeta^{-1*}, \qquad
	H_2^+ = \sqrt{\frac{5}{8}} \zeta^{+1} + \sqrt{\frac{3}{8}} \zeta^{-1*} \nonumber \\
	H_1^{++} &=& \sqrt{\frac{1}{4}} \zeta^{+2} + \sqrt{\frac{3}{4}} \zeta^{-2*}, \qquad
	H_2^{++} = -\sqrt{\frac{3}{4}} \zeta^{+2} + \sqrt{\frac{1}{4}} \zeta^{-2*} \nonumber \\
	H_1^{+3} &=& \sqrt{\frac{1}{8}} \zeta^{+3} - \sqrt{\frac{7}{8}} \zeta^{-3*}, \qquad
	H_2^{+3} = \sqrt{\frac{7}{8}} \zeta^{+3} + \sqrt{\frac{1}{8}} \zeta^{-3*}.
\end{eqnarray}
Again, all transitions among the mass eigenstates mediated by $W^{\pm}$ or $Z$ emission proceed with roughly comparable coupling strength.

\section{Constraints on couplings and masses}
\label{sec:constraints}

In this section we determine the constraints on the model parameters from perturbative unitarity, from the oblique parameters $S$, $T$, and $U$, and from the contributions of the new charged scalars to the loop-induced decays of the SM Higgs boson, $h \to \gamma\gamma$ and $h \to Z \gamma$.  We also comment on conditions to avoid alternate minima in which the $\zeta$ fields would have a nonzero vev.

Throughout this section, we show numerical results for the real neutral scalar in the mass range $m_{\zeta^{0,r}} = 80$--500~GeV.  We expect that scalars lighter than 80~GeV will be strongly constrained by searches at the CERN Large Electron-Positron (LEP) collider.  We leave the analysis of LEP constraints and CERN Large Hadron Collider (LHC) discovery prospects to future work.

\subsection{Unitarity constraints on scalar quartic couplings}
\label{sec:unitarityconstraints}

The scalar quartic couplings $\lambda_2$, $\lambda_3$ and $\lambda_4$ given in Eq.~(\ref{eq:conjugatepotential}) can be bounded by requiring perturbative unitarity of the zeroth partial wave scattering amplitudes.  The partial wave amplitudes are related to scattering matrix elements according to
\begin{equation}
	\mathcal{M} = 16\pi \sum_J (2J+1) a_J P_J(\cos\theta),
\end{equation}
where $J$ is the orbital angular momentum of the final state and $P_J(\cos\theta)$ is the corresponding Legendre polynomial.
Perturbative unitarity of the zeroth partial wave amplitude dictates the tree-level constraint,
\begin{equation}
	|{\rm Re} \, a_0 | \leq \frac{1}{2}.
	\label{rea0}
\end{equation}
We perform a coupled-channel analysis for processes $SS \to SS$ and $SS \to VV$, where $SS$ denotes any pair of scalars contained in $\Phi$ or $Z$ and $VV$ denotes any pair of transversely-polarized electroweak gauge bosons.  We include the $SS \to VV$ channels only for scalars contained in $Z$, whose amplitudes are enhanced by the large value of $n$.\footnote{This is the phenomenon that ultimately puts an upper limit on $n$~\cite{Hally:2012pu}.}
We work in the high-energy limit and treat the Goldstone bosons as physical particles in place of the longitudinal components of the electroweak gauge bosons.  For simplicity, we further neglect contributions from the quartic coupling $\lambda_1$, which is known to be small now that the SM-like Higgs boson mass has been measured, and from quartic couplings involving four $Z$ fields.  We find numerically that including such contributions leads to tighter constraints on $\lambda_2$, $\lambda_3$, and $\lambda_4$.  As such, our bounds are conservative. 

The scattering amplitudes are conveniently classified according to the total isospin and total hypercharge of the initial and final two-particle states.  The relevant amplitudes for the isospin-zero, hypercharge-zero channels are~\cite{Earl:2013jsa}
\begin{eqnarray}
	a_0([\zeta^*\zeta]_0 \to [\phi^*\phi]_0) &=& - \frac{\sqrt{n}}{8 \sqrt{2} \pi}\lambda_2, \nonumber \\
	a_0([\zeta^*\zeta]_0 \to [WW]_0) &=& \frac{g^2}{16 \pi} \frac{(n^2-1) \sqrt{n}}{2 \sqrt{3}}, \nonumber \\
	a_0([\zeta^*\zeta]_0 \to [BB]_0) &=& \frac{g^2}{16 \pi} \frac{s^2_W}{c^2_W} \frac{Y^2 \sqrt{n}}{2},
\end{eqnarray}
where the $\zeta^*\zeta \to WW,BB$ amplitudes include both of the contributing transverse gauge boson polarization combinations~\cite{Hally:2012pu}.  Here $g$ is the SU(2)$_L$ gauge coupling and $s_W, c_W \equiv \sin\theta_W, \cos\theta_W$ are the sine and cosine of the weak mixing angle, and $Y=1$ for our models.  We define the following normalized isospin-zero, hypercharge-zero field combinations,
\begin{eqnarray}
	[\phi^*\phi]_0 &=& \frac{1}{\sqrt{2}}(\phi^+\phi^- + \phi^{0*}\phi^0), \nonumber \\ {}
	[\zeta^*\zeta]_0 &=& \frac{1}{\sqrt{n}}\sum_{Q} \zeta^{Q*} \zeta^Q, \nonumber \\ {}
	[WW]_0 &=& \frac{1}{\sqrt{3}}\left(\sqrt{2} W^+W^- + \left(\frac{W^3 W^3}{\sqrt{2}}\right)\right),
		\nonumber \\ {}
	[BB]_0 &=& BB/\sqrt{2},
\end{eqnarray}
where the sum over $Q$ runs over the $n$ isospin eigenstates in $Z$ as shown in Eq.~(\ref{eq:states}).
The relevant amplitudes for the isospin-one, hypercharge-zero channels are~\cite{Earl:2013jsa}
\begin{eqnarray}
	a_0([\zeta^*\zeta]_1 \to [\phi^*\phi]_1) &=& - \frac{\sqrt{n(n^2-1)}}{32 \sqrt{6} \pi} \lambda_3, \nonumber \\
	a_0([\zeta^*\zeta]_1 \to [WB]_1) 
	&=& \frac{g^2}{16 \pi} \frac{s_W}{c_W} \frac{Y \sqrt{n (n^2 - 1)}}{\sqrt{6}},
\end{eqnarray}
where again the $\zeta^*\zeta \to WB$ amplitude includes both of the contributing transverse gauge boson polarization combinations~\cite{Hally:2012pu}.  Here we used the following normalized isospin-one, hypercharge-zero field combinations,
\begin{eqnarray}
	[\phi^*\phi]_1 &=& \frac{1}{\sqrt{2}}(\phi^+\phi^- - \phi^{0*}\phi^0), \nonumber \\ {}
	[\zeta^*\zeta]_1 &=& \sqrt{\frac{12}{n(n^2-1)}} \sum_{Q} \zeta^{Q*} T^3 \zeta^Q,
	\nonumber \\ {}
	[WB]_1 &=& W^3 B.
\end{eqnarray}
Finally, the relevant amplitude for the isospin-one, hypercharge-two channel is
\begin{equation}
	a_0([\zeta \zeta]_1 \to [\phi \phi]_1) = -\frac{\sqrt{n(n^2-1)}}{16\sqrt{6} \pi} \lambda_4,
\end{equation}
where the normalized isospin-one, hypercharge-two field combinations are
\begin{eqnarray}
	[\phi\phi]_1 &=& \phi^+ \phi^0, \nonumber \\ {}
	[\zeta \zeta]_1 &=& \sqrt{\frac{6}{n(n^2-1)}} (-1)^{n/2} 
	\sum_{j=1}^{n/2} (-1)^j (2j-1) \zeta^j \zeta^{-j+1}.
\end{eqnarray}

Finding the eigenvalues of the coupled-channel amplitude matrices and applying the constraint of Eq.~(\ref{rea0}), we obtain the following upper bounds on $|\lambda_2|$, $|\lambda_3|$ and $|\lambda_4|$:
\begin{eqnarray}
	|\lambda_2| &\leq& \sqrt{\frac{32 \pi^2}{n} - \frac{g^4 (n^2-1)^2}{24} - \frac{g^4 s_W^4}{8 c_W^4}}, \nonumber \\
	|\lambda_3| &\leq& 2 \sqrt{\frac{384 \pi^2}{n(n^2-1)} - g^4 \frac{s_W^2}{c_W^2}}, \nonumber \\
	|\lambda_4| &\leq& 8 \pi \sqrt{\frac{6}{n(n^2-1)}}.
\end{eqnarray}
Numerical values\footnote{We use $g^2 = 4 \pi \alpha/s_W^2$, $s_W^2 = 0.231$, and $\alpha = 1/128$.} for $n = 6$ and 8 are given in Table~\ref{lambdas}.

\begin{table}
\begin{center}
\begin{tabular}{l l l l}
\hline \hline
$n$~~ & $|\lambda_2|^{\rm max}$ & $|\lambda_3|^{\rm max}$ & $|\lambda_4|^{\rm max}$\\
\hline
6 & 6.59 & 8.48 & 4.25 \\
8 & 3.10 & 5.46 & 2.74 \\
\hline \hline
\end{tabular}
\end{center}
\caption{Upper limits on $|\lambda_2|$, $|\lambda_3|$ and $|\lambda_4|$ from perturbative unitarity for the models with $n = 6$ and 8.}
\label{lambdas}
\end{table}

\subsection{Constraints from the oblique parameters $S$, $T$ and $U$}
\label{sec:STUconstraints}

Further constraints can be obtained from experimental measurements of the electroweak oblique parameters $S$, $T$ and $U$~\cite{Peskin:1990zt}. These parameters probe new physics that can appear in loops in the electroweak gauge boson self-energies.  They were previously calculated for a large scalar multiplet with arbitrary isospin and hypercharge in the case of a global U(1) symmetry in Ref.~\cite{Lavoura:1993nq}.  We recomputed the contributions of a large scalar multiplet to $S$, $T$ and $U$ for the global $Z_2$-preserving case, in which the mass eigenstates do not always correspond to isospin eigenstates.  We checked that our results reduce to the U(1)-preserving limit when $\lambda_4 \to 0$.

For the $S$ parameter we find,
\begin{equation}
	S = \frac{s_W^2 c_W^2}{\pi}
	\sum_{i,j} \left(|C_{ijZ}|^2 - \frac{c_W^2-s_W^2}{s_W c_W} C_{ijZ}C_{ij\gamma}^*
		- |C_{ij\gamma}|^2\right) f_1(m_i,m_j),
	\label{eq:Sparam}
\end{equation}
where the couplings $C_{ijV}$ involving scalars $i$, $j$ and vector boson $V$ are defined with an overall factor of $e$ removed (see Appendix~\ref{sec:FRgauge}). The sums over states $i$ and $j$ run over $ij = \left\{ \zeta^{0,r}\zeta^{0,i},H_k^{+Q}H_l^{-Q},\zeta^{n/2}\zeta^{-n/2}\right\}$, where $kl=11, 12, 21, 22$ and $Q>0$.  
The dimensionless function $f_1(m_1,m_2)$ is defined as 

\begin{align}
	&f_1(m_1,m_2) = f_1(m_2,m_1) = \int_0^1 dx\,x(1-x)\log\left[x m_1^2 + (1-x)m_2^2\right]\,,\\
	& \quad= \left\{
	\begin{array}{l l}
	\frac{5(m_2^6-m_1^6)+27 (m_1^4 m_2^2-m_1^2 m_2^4)+12 (m_1^6-3 m_1^4 m_2^2) \log(m_1)+12(3 m_1^2 m_2^4-m_2^6)\log(m_2)}{36(m_1^2-m_2^2)^3} & {\rm for}\,\,m_1\neq m_2 \,,\\
	\frac{1}{6}\log m_1^2 & {\rm for}\,\,m_1= m_2\,.\\
	\end{array}
	\right.
\end{align}

For the $T$ parameter we find,
\begin{eqnarray}
	T &=& \frac{1}{4 \pi M_Z^2} \left[-\sum_r {S_r \left(\frac{C_{rr^*W^+W^-}}{c_W} - C_{rr^*ZZ}\right) f_2(m_r, m_r)}\right. \nonumber \\
	&& \left. \qquad\qquad - 2 \sum_{s, t} {\frac{\left|C_{stW^{+}}\right|^2}{c_W} f_2(m_s, m_t)} + 2 \sum_{i, j} {\left|C_{ijk}\right|^2 f_2(m_i, m_j)}\right]
	\label{eq:Tparam}
\end{eqnarray}
where the couplings $C_{rr^*XY}$ involving scalars $rr^*$ and vector bosons $X$, $Y$ are defined with an overall factor of $e^2$ removed (see Appendix~\ref{sec:FRgauge}). The sum over states $r$ runs over $r = \left\{ \zeta^{0,r}, \zeta^{0,i},  H_k^Q, \zeta^{n/2} \right\}$, with $Q > 0$ and $k = 1,2$.  For these couplings, $S_r$ is a symmetry factor given by $S_r=1/2$ for $r = \zeta^{0,r}$ or $\zeta^{0,i}$ and $S_r = 1$ otherwise.  The sums over states $s$ and $t$ run over $st = \left\{ \zeta^{0,r} H_k^{-}, \zeta^{0,i} H_k^{-}, H^{+Q}_k H^{-Q-1}_l, \zeta^{n/2} H_k^{-n/2-1} \right\}$, where again $kl=11, 12, 21, 22$ and $Q>0$.  The sums over states $i$ and $j$ run over the same set of states given below Eq.~(\ref{eq:Sparam}).  
The function $f_2(m_1,m_2)$ has dimensions of mass-squared and is defined as 
\begin{eqnarray}
	f_2(m_1,m_2) &=& f_2(m_2,m_1) 
	= \int_0^1 dx\, \left(x m_1^2+(1-x)m_2^2\right) \log\left[x m_1^2 + (1-x)m_2^2\right] 
	\nonumber \\
&	=& \left\{
	\begin{array}{l l}
	- \frac{1}{4} (m_1^2 + m_2^2)+ \frac{1}{m_1^2 - m_2^2}\left[m_1^4 \log m_1 - m_2^4 \log m_2\right] & {\rm for}\,\,m_1\neq m_2 \,,\\
	m_1^2 \log m_1^2 & {\rm for}\,\,m_1 =  m_2.
	\end{array}
	\right.
\end{eqnarray}

For the $U$ parameter we find,
\begin{equation}
	U = \frac{s_W^2}{\pi} \left[\sum_{s,t} |C_{stW^+}|^2 f_1(m_s,m_t) -\sum_{i,j} \left( c_W^2 |C_{ijZ}|^2 
	+ 2 s_W c_W C_{ijZ}C_{ij\gamma}^* + s_W^2 |C_{ij\gamma}|^2\right) f_1(m_i,m_j)\right]\,,
\end{equation}
where the sums over states $ij$ and $st$ run over the same sets of states given below Eqs.~(\ref{eq:Sparam}) and~(\ref{eq:Tparam}).

We impose a 95\% confidence level constraint from the oblique parameters using a $\chi^2$ built from the current experimental values, $S_{\rm exp}=0.03\pm0.10$, $T_{\rm exp}=0.05\pm0.12$, $U_{\rm exp}=0.03\pm0.10$, and the relative correlations $\rho_{ST}=0.89$, $\rho_{TU}=-0.83$, $\rho_{SU}=-0.54$~\cite{Baak:2012kk}.  The 95\% confidence level $\chi^2$ constraint in the 3-dimensional parameter space is given by\footnote{We use a $\chi^2$ constraint, rather than a $\Delta \chi^2 \equiv \chi^2 - \chi^2_{\rm min}$ relative to the minimum $\chi^2$ obtained in the model, so that parameter points that yield $S,T,U$ values too far from the experimental measurements will be excluded.  For comparison, the SM point $(S,T,U) = (0,0,0)$ has a $\chi^2$ value of 4.40 and the minimum $\chi^2$ obtained in the models is about 1.4 for $n=6$ and 1.3 for $n=8$.}
\begin{equation}
	\chi^2 = \sum_{i,j} (\mathcal{O}_i - \mathcal{O}_i^{\rm exp})
	(\mathcal{O}_j - \mathcal{O}_j^{\rm exp}) [\sigma^2]^{-1}_{ij}\leq 7.815,
\end{equation}
where $\mathcal{O}_i$ is the $i$th observable of the set ${S,T,U}$ and $[\sigma^{2}]^{-1}_{ij}$ is the inverse of the matrix of uncertainties,
\begin{equation}
	[\sigma^2]_{ij} = \Delta \mathcal{O}_i \, \Delta \mathcal{O}_j \, \rho_{ij},
\end{equation}
where $\rho_{ij}$ are the relative correlations (note that $\rho_{ii} \equiv 1$).

\begin{figure}
\begin{center}
\resizebox{0.49\textwidth}{!}{
\includegraphics{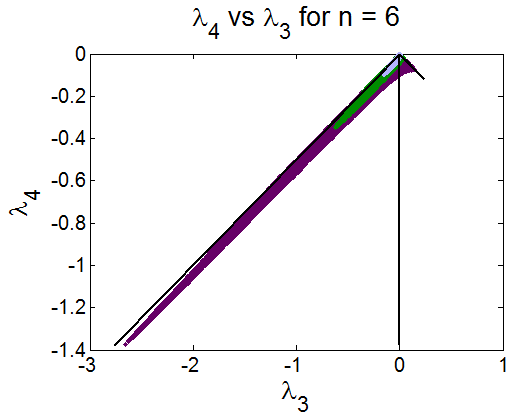}}
\resizebox{0.49\textwidth}{!}{
\includegraphics{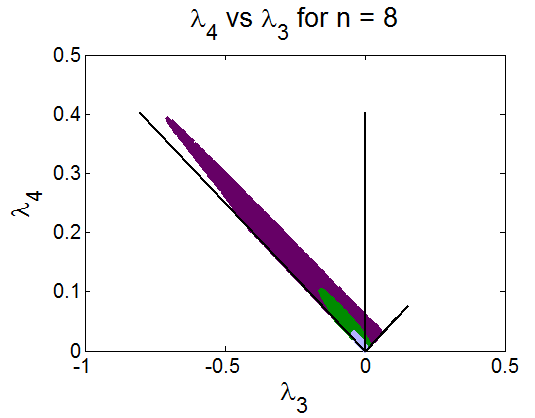}}
\caption{Values of $\lambda_3$ and $\lambda_4$ allowed at 95\% confidence level by the oblique parameters, under the requirement that $|\lambda_3| \leq 2|\lambda_4|$, for $m_{\zeta^{0,r}} = 80$, 150, and 300~GeV (light blue, medium green, and dark purple, respectively).  Left: $n = 6$ model.  Right: $n = 8$ model.}
\label{ZL3L4}
\end{center}
\end{figure}

The contributions of the large multiplet to $S$, $T$ and $U$ depend only on the mass spectrum and mass-eigenstate compositions.  Thus the oblique parameters constrain only $\lambda_3$, $\lambda_4$, and the overall mass scale, which can be parameterized by $m_{\zeta^{0,r}}$.
The ranges of $\lambda_3$ and $\lambda_4$ allowed by the oblique parameters for sample values $m_{\zeta^{0,r}} = 80$, 150, and 300~GeV are shown in Fig.~\ref{ZL3L4}.  We have imposed the constraint $|\lambda_3| \leq 2 |\lambda_4|$, indicated by the diagonal lines, which is required for $\zeta^{0,r}$ to be the lightest state.

As can be seen in Fig.~\ref{ZL3L4}, the allowed parameter space is tightly constrained to lie near the $|\lambda_3| = 2 |\lambda_4|$ line.  This is driven by the contributions of the large multiplet to the $T$ and $U$ parameters, which rapidly become large away from this line.  This feature leads to the clustering of the mass eigenstates into two groups as discussed in Sec.~\ref{sec:models}.  Furthermore, unless both $\lambda_3$ and $\lambda_4$ are very small, the oblique parameters require $\lambda_3 < 0$.  This results in the state with the largest electric charge ($\zeta^{+3}$ for $n=6$ and $\zeta^{+4}$ for $n=8$) to be clustered with the heavier group of mass eigenstates.  The length of the allowed region along the $|\lambda_3| \simeq 2 |\lambda_4|$ line is limited mainly by the $S$ parameter constraint.

\begin{figure}
\begin{center}
\resizebox{0.49\textwidth}{!}{
\includegraphics{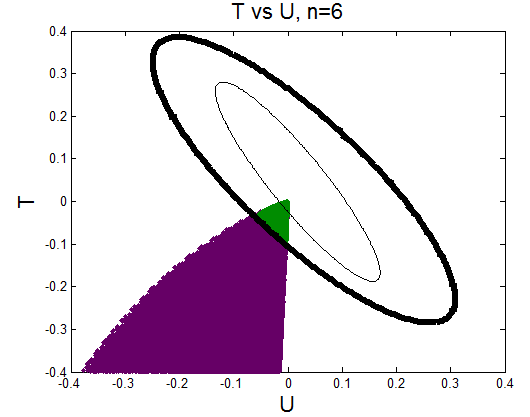}
}
\resizebox{0.49\textwidth}{!}{
\includegraphics{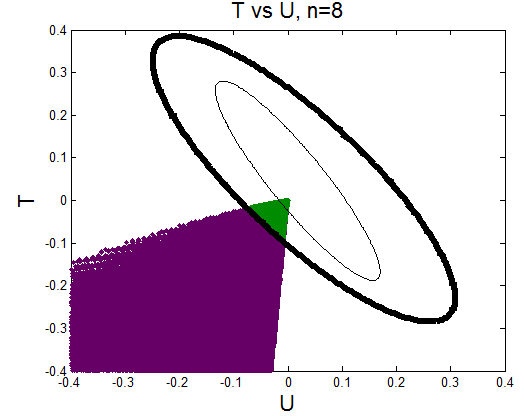}}
\caption{Unitarity-allowed $U$ and $T$ values for $m_{\zeta^{0,r}} = 80$--500~GeV (shaded regions) and the projection of the $S,T,U$ 95\% confidence level constraint ellipsoid (thick black ellipse).  Points allowed by the $S,T,U$ constraint are shown in green (lighter shaded region). The thin black ellipse indicates the slice through the three-dimensional constraint ellipsoid at $S=0$.  Left: $n=6$ model.  Right: $n=8$ model.}
\label{UvsT}
\end{center}
\end{figure}

\begin{figure}
\begin{center}
\resizebox{0.49\textwidth}{!}{
\includegraphics{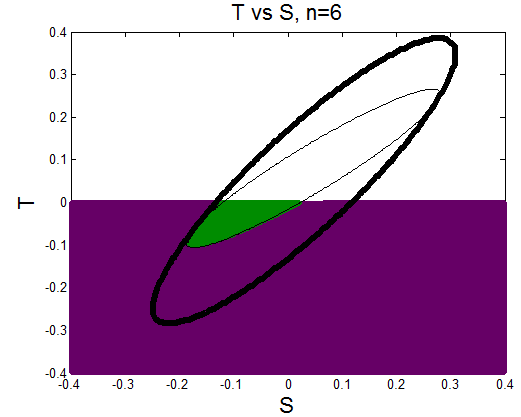}
}
\resizebox{0.49\textwidth}{!}{
\includegraphics{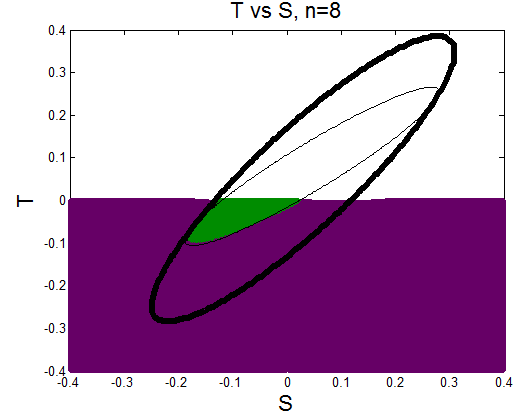}}
\caption{Unitarity-allowed $S$ and $T$ values for $m_{\zeta^{0,r}} = 80$--500~GeV (shaded regions) and the projection of the $S,T,U$ 95\% confidence level constraint ellipsoid (thick black ellipse). Points allowed by the $S,T,U$ constraint are shown in green (lighter shaded region). The thin black ellipse indicates the slice through the three-dimensional constraint ellipsoid at $U=0$.  Left: $n=6$ model.  Right: $n=8$ model.}
\label{TvsS}
\end{center}
\end{figure}

\begin{figure}
\begin{center}
\resizebox{0.49\textwidth}{!}{
\includegraphics{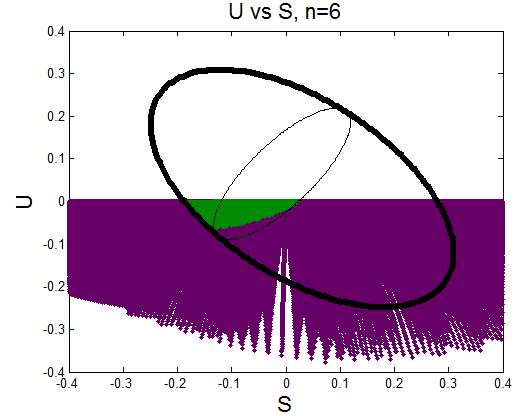}
}
\resizebox{0.49\textwidth}{!}{
\includegraphics{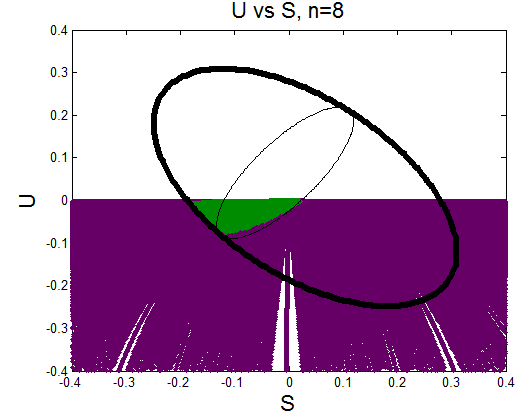}}
\caption{Unitarity-allowed $S$ and $U$ values for $m_{\zeta^{0,r}} = 80$--500~GeV (shaded regions) and the projection of the $S,T,U$ 95\% confidence level constraint ellipsoid (thick black ellipse). Points allowed by the $S,T,U$ constraint are shown in green (lighter shaded region). The thin black ellipse indicates the slice through the three-dimensional constraint ellipsoid at $T=0$. Scatter in the plot is due to the numerical scan.  Left: $n=6$ model.  Right: $n=8$ model.}
\label{UvsS}
\end{center}
\end{figure}

To better understand the source of these constraints, we plot the projections of the $S,T,U$ 95\% confidence level constraint ellipsoid, along with the region populated by the models, in Figs.~\ref{UvsT}, \ref{TvsS}, and~\ref{UvsS}.  The $T$ and $U$ measurements severely constrain the parameter space, as can be seen in Fig.~\ref{UvsT}.  This in turn tightens the allowed excursion of the $S$ parameter, as shown in Figs.~\ref{TvsS} and~\ref{UvsS}.  In particular, because of the correlations among the measured values of $S$, $T$, and $U$, positive values of $S$ are severely constrained.  The sign of $S$ is the same as the sign of $\lambda_3$ in our models.  This constraint therefore leads to the preference for negative values of $\lambda_3$.

\begin{figure}
\begin{center}
\resizebox{0.49\textwidth}{!}{
\includegraphics{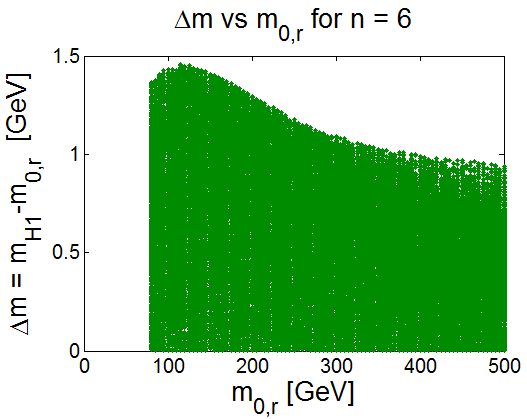}
}
\resizebox{0.49\textwidth}{!}{
\includegraphics{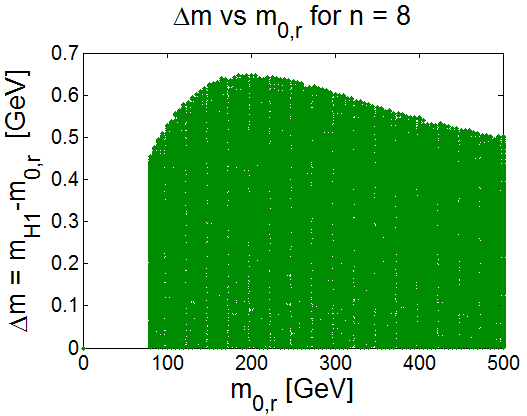}}
\caption{Allowed range for the first mass splitting $\Delta m \equiv m_{H_1^+} - m_{\zeta^{0,r}}$ as a function of $m_{\zeta^{0,r}}$.  Left: $n=6$ model.  Right: $n=8$ model.}
\label{DM1}
\end{center}
\end{figure}

The oblique parameters tightly constrain the mass splittings among the lightest states of the large multiplet.  The first mass splitting between $\zeta^{0,r}$ and $H_1^+$ must be quite small: for $m_{\zeta^{0,r}} \leq 500$~GeV, this splitting $\Delta m \equiv m_{H_1^+} - m_{\zeta^{0,r}}$ is less than 1.5~GeV in the $n=6$ model and less than 0.7~GeV in the $n=8$ model (see Fig.~\ref{DM1}).

\begin{figure}
\begin{center}
\resizebox{0.49\textwidth}{!}{
\includegraphics{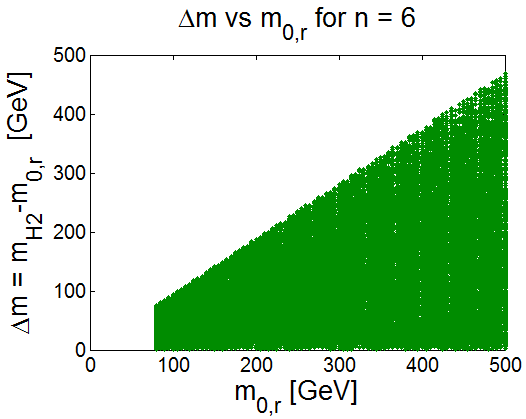}
}
\resizebox{0.49\textwidth}{!}{
\includegraphics{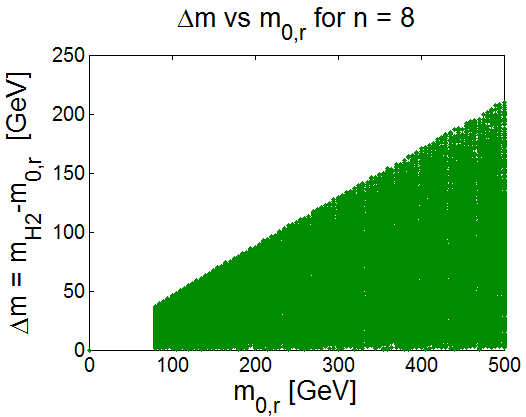}}
\caption{Allowed range for one of the larger mass splittings, $\Delta M \equiv m_{H_2^+} - m_{\zeta^{0,r}}$, as a function of $m_{\zeta^{0,r}}$.  Left: $n=6$ model.  Right: $n=8$ model.}
\label{DM2}
\end{center}
\end{figure}

The splitting between the heavier group of states and the lighter group of states, on the other hand, is much less tightly constrained.  The allowed mass splitting between the lightest state $\zeta^{0,r}$ and the heavier singly-charged state $H_2^+$, $\Delta M \equiv m_{H_2^+} - m_{\zeta^{0,r}}$, is shown in Fig.~\ref{DM2} as a function of $m_{\zeta^{0,r}}$.  For the $n=6$ model, this splitting can be as large as 100~GeV for $m_{\zeta^{0,r}} = 100$~GeV, growing to 450~GeV for $m_{\zeta^{0,r}} = 500$~GeV.  For the $n=8$ model the maximum allowed mass splitting is about half as large, ranging between about 50~GeV and about 200~GeV for $m_{\zeta^{0,r}}$ between 100 and 500~GeV.  The linear growth of the maximum allowed mass splitting shown in Fig.~\ref{DM2} is eventually cut off by the unitarity constraints on $\lambda_3$ and $\lambda_4$ for large enough $m_{\zeta^{0,r}}$.

\subsection{Constraints from the loop-induced Higgs decay $h \to \gamma\gamma$}
\label{sec:hgaga}

To constrain the parameter $\lambda_2$, we consider the one-loop contributions of the charged scalars to the Higgs decay $h\rightarrow\gamma\gamma$.\footnote{Contributions to $h \to \gamma\gamma$ from multiply-charged scalars in the loop have also been considered in Refs.~\cite{Cai:2012kt,Chen:2012vm}.}  
The decay partial width for $h \to \gamma\gamma$ can be written as~\cite{HHG}
\begin{equation}
	\Gamma(h\rightarrow\gamma\gamma) = 
	\frac{\alpha^2 g^2}{1024\pi^3}\frac{M_h^3}{M_W^2}\left|\sum_i N_{ci} Q_i^2 F_i\right|^2,
\end{equation}
where $i$ sums over charged particles of spin $0$, 1/2, and 1, $Q_i$ is the electric charge of the particle in the loop in units of $e$, $N_{ci}$ is the color multiplicity (3 for quarks and 1 for color-singlet particles) and $F_i$ is a function that depends on the spin and mass of the particle in the loop~\cite{HHG}:
\begin{eqnarray}
	F_1 &=& 2+3\tau+3\tau(2-\tau) f(\tau) \nonumber \\
	F_{1/2} &=& -2\tau[1+(1-\tau) f(\tau)] \nonumber \\
	F_0 &=& \beta\tau[1-\tau f(\tau)].
\end{eqnarray}
Here $\tau_i = 4 m_i^2/M^2_h$, where $m_i$ is the mass of the particle in the loop, and the function $f(\tau)$ is given by~\cite{HHG}
\begin{equation}
	f(\tau) = \left\{ \begin{array}{c c}
	\left[\arcsin\left(\sqrt{\frac{1}{\tau}}\right)\right]^2 &\quad  {\rm if}\,\,\tau\geq 1, \\
	-\frac{1}{4}\left[\log\left(\frac{\eta_+}{\eta_-}\right)-i\,\pi\right]^2 &\quad  {\rm if}\,\,\tau< 1,
	\end{array} \right.
	\label{eq:ftau}
\end{equation}
with $\eta_{\pm} \equiv 1 \pm \sqrt{1-\tau}$.
In the numerical computation of partial widths we include only the top quark and $W$ boson contributions, as well as the new scalars; contributions from the lighter fermions are small. 

For a scalar particle in the loop, we have inserted the factor $\beta$ into the definition of $F_0$ to capture the coupling of the scalar to the Higgs,
\begin{equation}
	\beta = C_{hss} \frac{M_W}{g m_s^2} = C_{hss} \frac{v}{2 m_s^2}.
\end{equation}
The couplings $C_{hss}$ for our models are collected in Appendix~\ref{sec:rulesB}.  Note that all couplings of the Higgs to pairs of scalars from the large multiplet are diagonal in the mass basis.  The functions $\beta$ for the scalars $H^Q_i$ and $\zeta^{+n/2}$ in the loop can be written as
\begin{eqnarray}
	\beta_{H_{1,2}^{Q}} &=& \frac{v^2}{2 m_{H_{1,2}^Q}^2}\left[\lambda_2+\frac{\lambda_3}{4}\mp\frac{1}{2}\sqrt{Q^2\lambda_3^2+(n^2-4Q^2)\lambda_4^2}\right], \nonumber \\
	\beta_{\zeta^{+n/2}} &=& \frac{v^2}{2 m_{\zeta^{+n/2}}^2}\left[\lambda_2-\frac{n-1}{4}\lambda_3\right].
\end{eqnarray}
Because these couplings depend on $\lambda_2$, measurements of the Higgs decay $h \to \gamma\gamma$ can be used to put additional constraints on $\lambda_2$ as a function of $\lambda_3$, $\lambda_4$, and $m_{\zeta^{0,r}}$.  The ATLAS and CMS experiments have measured the Higgs signal strength $\mu_{\gamma\gamma}$ in the $\gamma\gamma$ final state, defined relative to the SM prediction.  Because Higgs production rates are not modified in our models, and because the only significant effect of the new scalars on Higgs decays is through modification of the partial widths of the rare loop-induced processes $h \to \gamma\gamma$ and $h \to Z \gamma$, we have to a very good approximation 
\begin{equation}
	\mu_{\gamma\gamma} \simeq R_{\gamma\gamma} \equiv \frac{\Gamma(h \to \gamma\gamma)}{\Gamma_{\rm SM}(h \to \gamma\gamma)},
	\label{eq:Rgaga}
\end{equation}
and an analogous expression for $R_{Z\gamma}$.
The measured values of this rate from ATLAS and CMS are summarized in Table~\ref{hggmeas}.

\begin{table}
\begin{center}
\renewcommand{\arraystretch}{1.5}
\begin{tabular}{c c c}
\hline \hline
 Observable & ATLAS & CMS \\
\hline
$\mu_{\gamma\gamma}$ & $1.65\stackrel{+0.32}{_{-0.30}}$~\cite{hgaga-ATLAS} & $0.78\stackrel{+0.28}{_{-0.26}}$~\cite{hgaga-CMS} \\
$\mu_{Z \gamma}$  & $<18.2$~\cite{hgz-ATLAS} & $<10$~\cite{hgz-CMS} \\
\hline \hline
\end{tabular}
\end{center}
\caption{Current measurements of the signal strengths for $h\rightarrow\gamma\gamma$ and $h\rightarrow Z \gamma$ relative to the SM predictions.  For $h \to Z \gamma$ we quote the 95\% confidence level upper bounds for $M_h = 125$~GeV.}
\label{hggmeas}
\end{table}

\begin{figure}
\begin{center}
\resizebox{0.49\textwidth}{!}{
\includegraphics{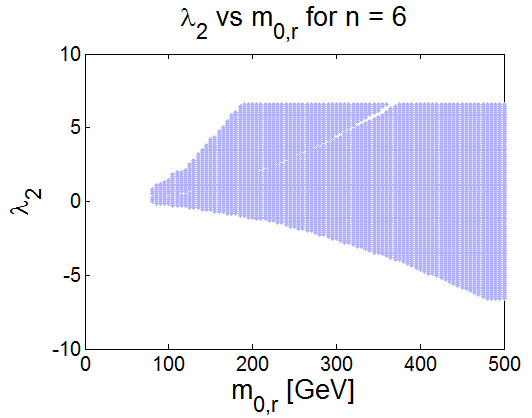}
}
\resizebox{0.49\textwidth}{!}{
\includegraphics{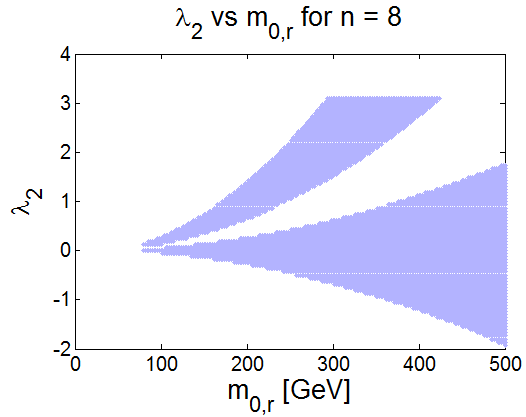}}
\caption{Allowed ranges of $\lambda_2$ as a function of $m_{\zeta^{0,r}}$ after imposing the LHC constraint on the Higgs decay to two photons, together with the oblique parameter constraints and perturbative unitarity.  Left: $n=6$ model.  Right: $n=8$ model.}
\label{L2vsM0}
\end{center}
\end{figure}

We find the allowed range of $\lambda_2$ as a function of $m_{\zeta^{0,r}}$ by scanning over the values of $\lambda_3$ and $\lambda_4$ allowed by the oblique parameter constraints and perturbative unitarity.  We accept points for which $R_{\gamma\gamma}$ falls within the $2\sigma$ range for $\mu_{\gamma\gamma}$ of either the ATLAS or CMS measurement.  Results are shown in Fig.~\ref{L2vsM0}.  Note that the constraint from perturbative unitarity, $|\lambda_2| \leq 6.59$ (3.10) for $n = 6$ (8), is visible in the plots.  The upper branch of allowed $\lambda_2$ values, clearly visible in the right panel of Fig.~\ref{L2vsM0} for $n=8$, corresponds to a sign flip of the total $h \to \gamma\gamma$ amplitude relative to the SM prediction.  This is separated from the rest of the points due to the lower bound on $\mu_{\gamma\gamma}$.  The same feature is present in the $n=6$ case, though it is less clearly visible in Fig.~\ref{L2vsM0} due to the wider allowed ranges of $\lambda_3$ and $\lambda_4$.

The application of the $h\rightarrow\gamma\gamma$ constraint does not significantly restrict the range of $\lambda_3$ or $\lambda_4$ beyond the constraints already obtained from the oblique parameters and perturbative unitarity. 

\subsection{Predictions for $h \to Z \gamma$}

The charged scalars in our models also contribute to the loop-induced decay $h \to Z \gamma$.  The decay partial width for this process can be written as (see, e.g., Ref.~\cite{HHG})
\begin{equation}
	\Gamma(h\rightarrow Z\gamma) = 
	\frac{\alpha^2}{512\pi^3}\left|\frac{2}{v}\left(A_F + A_W\right)+A_S\right|^2 
	M_h^3 \left[1-\frac{M_Z^2}{M_h^2}\right]^3,
\end{equation}
where the contributions to the amplitude from fermions, the $W$ boson, and scalars are given by\footnote{It was pointed out in Ref.~\cite{HGZ-Chen} that there are some discrepancies in the literature among different calculations of the contributions of charged scalars to the amplitude for $h\rightarrow Z \gamma$.  Our formulas are consistent with those of Refs.~\cite{HHG,HGZ-Chen,HGZ-Carena}.}
\begin{eqnarray}
	A_F &=& \sum_f N_{cf} \frac{-2 Q_f \left(T^{3L}_f - 2 Q_f s_W^2 \right)}{s_W c_W}
	\left[I_1(\tau_f,\lambda_f) - I_2(\tau_f,\lambda_f)\right], \nonumber \\
	A_W &=& -\cot\theta_W\left\{4\left(3-\tan^2\theta_W\right) I_2\left(\tau_W,\lambda_W\right)+\left[\left(1+\frac{2}{\tau_W}\right)\tan^2\theta_W-\left(5+\frac{2}{\tau_W}\right)\right] I_1\left(\tau_W,\lambda_W\right)\right\}, \nonumber \\
	A_S &=& 2\sum_s\frac{C_{hss} C_{ssZ} Q_s}{m_s^2} I_1(\tau_s, \lambda_s),
\end{eqnarray}
where the scalar couplings $C_{hss}$ and $C_{ssZ}$ are given in Appendix~\ref{sec:rulesB} and $\lambda_i$ is defined analogously to $\tau_i$ but with $M_h$ replaced by $M_Z$:
\begin{equation}
	\lambda_i = \frac{4 m_i^2}{M_Z^2}.
\end{equation}
The sums over $f$ and $s$ run over all charged fermions and scalars in the model (in our numerical calculation, we neglect all fermions other than the top quark).

The functions $I_i(a,b)$ are defined as~\cite{HHG}
\begin{eqnarray}
	I_1(a,b) &=& \frac{ab}{2(a-b)} + \frac{a^2b^2}{2 (a-b)^2}\left[f(a)-f(b)\right]
	+ \frac{a^2b}{(a-b)^2} \left[g(a) - g(b)\right], \nonumber \\
	I_2(a,b) &=& -\frac{a b}{2 (a-b)}\left[f(a)-f(b)\right],
\end{eqnarray}
where the function $f(\tau)$ was given in Eq.~(\ref{eq:ftau}) and the function $g(\tau)$ is defined as 
\begin{equation}
	g(\tau) = \left\{ \begin{array}{c c}
	\sqrt{\tau-1}\arcsin\left(\sqrt{\frac{1}{\tau}}\right) &\quad  {\rm if}\,\,\tau\geq 1, \\
	\frac{1}{2}\sqrt{1-\tau}\left[\log\left(\frac{\eta_+}{\eta_-}\right)-i\,\pi\right] &\quad  {\rm if}\,\,\tau< 1.
	\end{array} \right.
\end{equation}

\begin{figure}
\begin{center}
\resizebox{0.5\textwidth}{!}{
\includegraphics{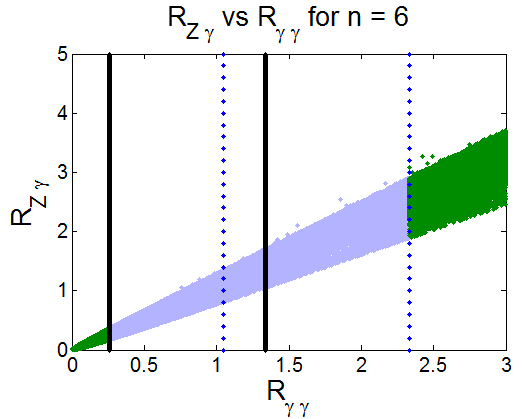}
}
\resizebox{0.49\textwidth}{!}{
\includegraphics{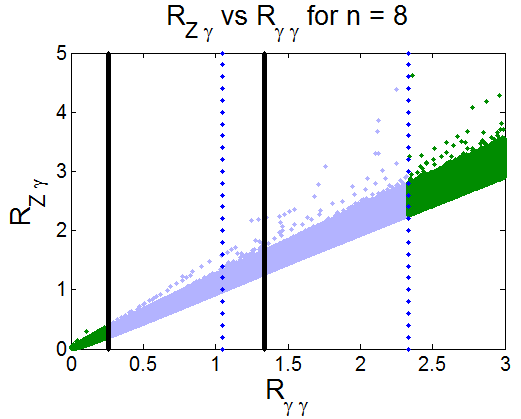}}
\caption{Allowed range of the $h \to Z\gamma$ partial width scaling factor $R_{Z\gamma}$ plotted against the $h \to\gamma\gamma$ partial width scaling factor $R_{\gamma\gamma}$ [see Eq.~(\ref{eq:Rgaga})].  Colored points represent a scan over the parameter space allowed by perturbative unitarity and the oblique parameters $S$, $T$, and $U$, for $m_{\zeta^{0,r}}$ in the range 80--500~GeV.  The light blue points are also allowed at $2\sigma$ by the experimental measurement of the Higgs signal strength in two photons from either ATLAS~\cite{hgaga-ATLAS} (dotted vertical lines) or CMS~\cite{hgaga-CMS} (solid vertical lines).  Left: $n=6$ model.  Right: $n=8$ model.}
\label{HGZvsHGG}
\end{center}
\end{figure}

The LHC experiments currently constrain the rate for $h \to Z\gamma$ to be less than about 10 times the SM prediction (see Table~\ref{hggmeas}).  This does not further constrain the parameter space of our models once the constraints from perturbative unitarity, oblique parameters, and $h \to \gamma\gamma$ have been applied.  Indeed, we find $R_{Z\gamma}$ to be strongly correlated with $R_{\gamma\gamma}$, as shown in Fig.~\ref{HGZvsHGG}.  Imposing all experimental constraints previously discussed and scanning $m_{\zeta^{0,r}}$ in the range 80--500~GeV, we find that the ratio $R_{Z\gamma}/R_{\gamma\gamma}$ lies between approximately 0.7 and 1.3 for the $n=6$ model, and between approximately 0.8 and 1.2 for the $n=8$ model but with some points higher than the quoted range, as can be seen in Fig.~\ref{HGZvsHGG}.

\subsection{Condition to avoid $Z_2$-breaking minima in the scalar potential}

In any extended Higgs sector, the avoidance of alternative minima in the scalar potential---which could potentially break electric charge or other desired symmetries of the model---can provide additional constraints on the model parameters.  In our models, a sufficient\footnote{We have not proved that this condition is necessary.} condition to avoid alternate minima in which one or more of the $\zeta$ fields acquires a vev is to require $M^2 > 0$ in Eq.~(\ref{eq:conjugatepotential}).  This holds so long as the potential is bounded from below and none of the fields are tachyonic at the desired electroweak symmetry breaking vacuum. For a fixed value of $m_{\zeta^{0,r}}$, this imposes an upper limit on $\Lambda_6$ or $\Lambda_8$ via Eqs.~(\ref{eq:lambda6}) or (\ref{eq:lambda8}), respectively.  As we will see in the next sections, this constraint has an effect on the predictions for the allowed thermal relic density of $\zeta^{0,r}$ and its direct detection cross section.

\section{Thermal relic density}
\label{sec:relic}

The relic abundance of $\zeta^{0,r}$ is determined by its interactions in the early Universe. If we assume a standard thermal history---i.e., that the temperature was high enough at one time for $\zeta^{0,r}$ to have been in thermal equilibrium, and that no late-decaying relics enhanced or diluted the $\zeta^{0,r}$ density---then the relic density of $\zeta^{0,r}$ at the present time can be computed from its annihilation rate in the early universe.  For generic relics, the density will be inversely proportional to the annihilation cross-section, $\Omega_X \propto \langle\sigma_X v_{rel} \rangle^{-1}$~\cite{Steigman:2012nb}, where $v_{rel}$ is the relative velocity of the two particles in the annihilation collision and the brackets indicate an average over this velocity distribution at the time of freeze-out.  Such an average is numerically necessary only if the annihilation cross section vanishes in the $v_{rel} \to 0$ limit.  Because of this simple relationship, we can determine the fraction of the total dark matter that is made up by $\zeta^{0,r}$ using the formula
\begin{equation}
	\frac{\Omega_{\zeta^{0,r}}}{\Omega_{\rm DM}} 
	= \frac{\langle\sigma v_{rel} \rangle_{\rm std}}
	{\langle\sigma v_{rel}(\zeta^{0,r}\zeta^{0,r} \to {\rm any})\rangle},
	\label{eq:relicratio}
\end{equation}
where $\Omega_{\rm DM}$ is the current total dark matter relic abundance and $\langle \sigma v_{rel}\rangle_{\rm std}$ is the ``standard'' annihilation cross section required to obtain this total dark matter relic abundance, for which we use $\langle \sigma v_{rel}\rangle_{\rm std} = 3 \times 10^{-26}$~cm$^3/$s~\cite{Steigman:2012nb}.

\subsection{Annihilations to two-body final states}

The $\zeta^{0,r}$ is a self-annihilating particle which interacts with the SM via gauge or Higgs boson exchange. As such, the final states for the annihilation of two $\zeta^{0,r}$ particles include $W^{+}W^{-}$, $ZZ$, $hh$, and $f \bar f$ (via $s$-channel Higgs exchange).  We neglect co-annihilations with other scalars from the $Z_2$-odd multiplet.  We compute the annihilation cross sections in the zero-velocity limit. Because these cross sections are all nonzero in this limit, we do not need to average over the velocity distribution.

The annihilation cross sections to two-body final states are given in the $v_{rel} \to 0$ limit by,
\begin{eqnarray}
	\sigma v_{rel}(\zeta^{0,r} \zeta^{0,r} \to W^+ W^-) &=& 
	\frac{M_W^4}{8 \pi v^4}
	\sqrt{1 - \frac{M_W^2}{m_{\zeta^{0,r}}^2}} \left[
	\frac{A_{W}^2}{m_{\zeta^{0,r}}^2} \left( 3 - 4 \frac{m_{\zeta^{0,r}}^2}{M_W^2} + 4 \frac{m_{\zeta^{0,r}}^4}{M_W^4} \right) \right. \nonumber \\
	&& \left. + 2 A_{W} B_{W} \left( 1 - 3 \frac{m_{\zeta^{0,r}}^2}{M_W^2} + 2 \frac{m_{\zeta^{0,r}}^4}{M_W^4} \right) + B_{W}^2 m_{\zeta^{0,r}}^2 \left( 1 - \frac{m_{\zeta^{0,r}}^2}{M_W^2}\right)^2 \right], \nonumber \\
	\sigma v_{rel}(\zeta^{0,r} \zeta^{0,r} \to Z Z) &=& 
	\frac{M_Z^4}{16 \pi v^4}
	\sqrt{1 - \frac{M_Z^2}{m_{\zeta^{0,r}}^2}} \left[
	\frac{A_Z^2}{m_{\zeta^{0,r}}^2} \left( 3 - 4 \frac{m_{\zeta^{0,r}}^2}{M_Z^2} + 4 \frac{m_{\zeta^{0,r}}^4}{M_Z^4} \right) \right. \nonumber \\
	&& \left. + 2 A_Z B_Z \left( 1 - 3 \frac{m_{\zeta^{0,r}}^2}{M_Z^2} 
	+ 2 \frac{m_{\zeta^{0,r}}^4}{M_Z^4} \right) 
	+ B_Z^2 m_{\zeta^{0,r}}^2 \left(1 - \frac{m_{\zeta^{0,r}}^2}{M_Z^2}\right)^2 \right], \nonumber \\
	\sigma v_{rel}(\zeta^{0,r} \zeta^{0,r} \to hh) &=& \frac{\Lambda_n^2}{64\pi m_{\zeta^{0,r}}^2}
	\sqrt{1 - \frac{M_h^2}{m_{\zeta^{0,r}}^2}}
	\left[1 + \frac{3 M_h^2}{4m_{\zeta^{0,r}}^2 - M_h^2} 
	- \frac{2v^2\Lambda_n}{2m_{\zeta^{0,r}}^2 - M_h^2} \right]^2, \nonumber \\
	\sigma v_{rel}(\zeta^{0,r} \zeta^{0,r} \to f \bar f) &=& \frac{N_c}{4\pi} 
	\left[1 - \frac{m_f^2}{m_{\zeta^{0,r}}^2}\right]^{3/2}
	\frac{m_f^2 \Lambda_n^2}{(4m_{\zeta^{0,r}}^2 - M_h^2)^2},
	\label{eq:annihilation}
\end{eqnarray}
where $n=6, 8$ is the size of the multiplet, $N_c$ is the number of colors of the final-state fermions, and $v \simeq 246$~GeV is the usual Higgs vacuum expectation value. The coefficients used in the cross section formulas for $\zeta^{0,r}\zeta^{0,r} \to W^+W^-$ and $\zeta^{0,r}\zeta^{0,r} \to ZZ$ are given by
\begin{eqnarray}
	A_Z &=& 1 + \frac{\Lambda_n v^2}{4 m_{\zeta^{0,r}}^2 - M_h^2}\\
	B_Z &=& \frac{4}{M_Z^2 - m_{\zeta^{0,r}}^2 - m_{\zeta^{0,i}}^2}\\
	A_W &=& \frac{n^2 - 2}{2} + \frac{\Lambda_n v^2}{4 m_{\zeta^{0,r}}^2 - M_h^2}\;,\\
	B_W &=& \frac{\left(n \cos\alpha_1 - \sqrt{n^2 - 4}\, \sin\alpha_1\right)^2}
	{M_W^2 - m_{\zeta^{0,r}}^2 - m_{H_1^{+}}^2} 
	+ \frac{\left(-n \sin\alpha_1 - \sqrt{n^2 - 4}\, \cos\alpha_1\right)^2}
	{M_W^2 - m_{\zeta^{0,r}}^2 - m_{H_2^{+}}^2}.
\end{eqnarray}
The combinations of couplings $\Lambda_6$ and $\Lambda_8$ were defined in Eqs.~(\ref{eq:lambda6}) and (\ref{eq:lambda8}), respectively; they can both be expressed by the formula
\begin{equation}
	\Lambda_n = \lambda_2 + \frac{1}{4}\lambda_3 + \frac{n}{2}(-1)^{n/2+1}\lambda_4.
	\label{eq:lambdan}
\end{equation}

We note that when $m_{\zeta^{0,r}} \gg M_W,M_Z$, the annihilation cross sections for $\zeta^{0,r} \zeta^{0,r} \to W^+W^-$ and $ZZ$ go like $1/m_{\zeta^{0,r}}^2$.  In this limit, the new scalars become increasingly degenerate due to the constraints on the size of $|\lambda_3|$ and $|\lambda_4|$; the values of $A_{W,Z}$ and $B_{W,Z}$ are then related in such a way as to allow a cancellation of the $m_{\zeta^{0,r}}^2/M_{W,Z}^2$ and $m_{\zeta^{0,r}}^4/M_{W,Z}^4$ terms in the square brackets, which would otherwise make the cross section grow with increasing $m_{\zeta^{0,r}}$.  This cancellation provides a nice cross-check of the matrix element calculation.  We also checked our analytic results using CalcHEP~\cite{Belyaev:2012qa}.

\subsection{Numerical results}
\label{sec:omegaplots}

\begin{figure}
\begin{center}
\includegraphics[width=0.49\textwidth]{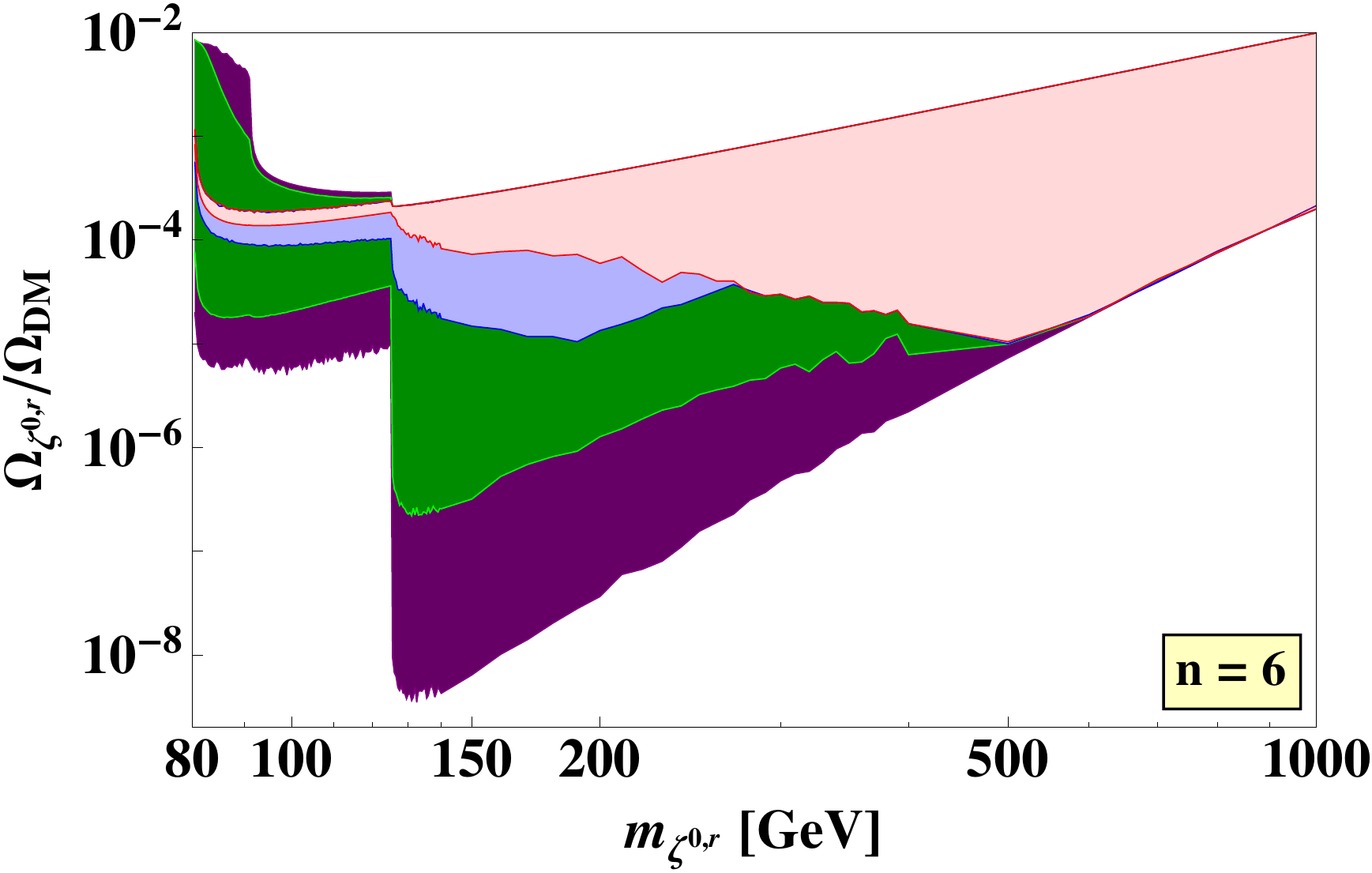}
\includegraphics[width=0.49\textwidth]{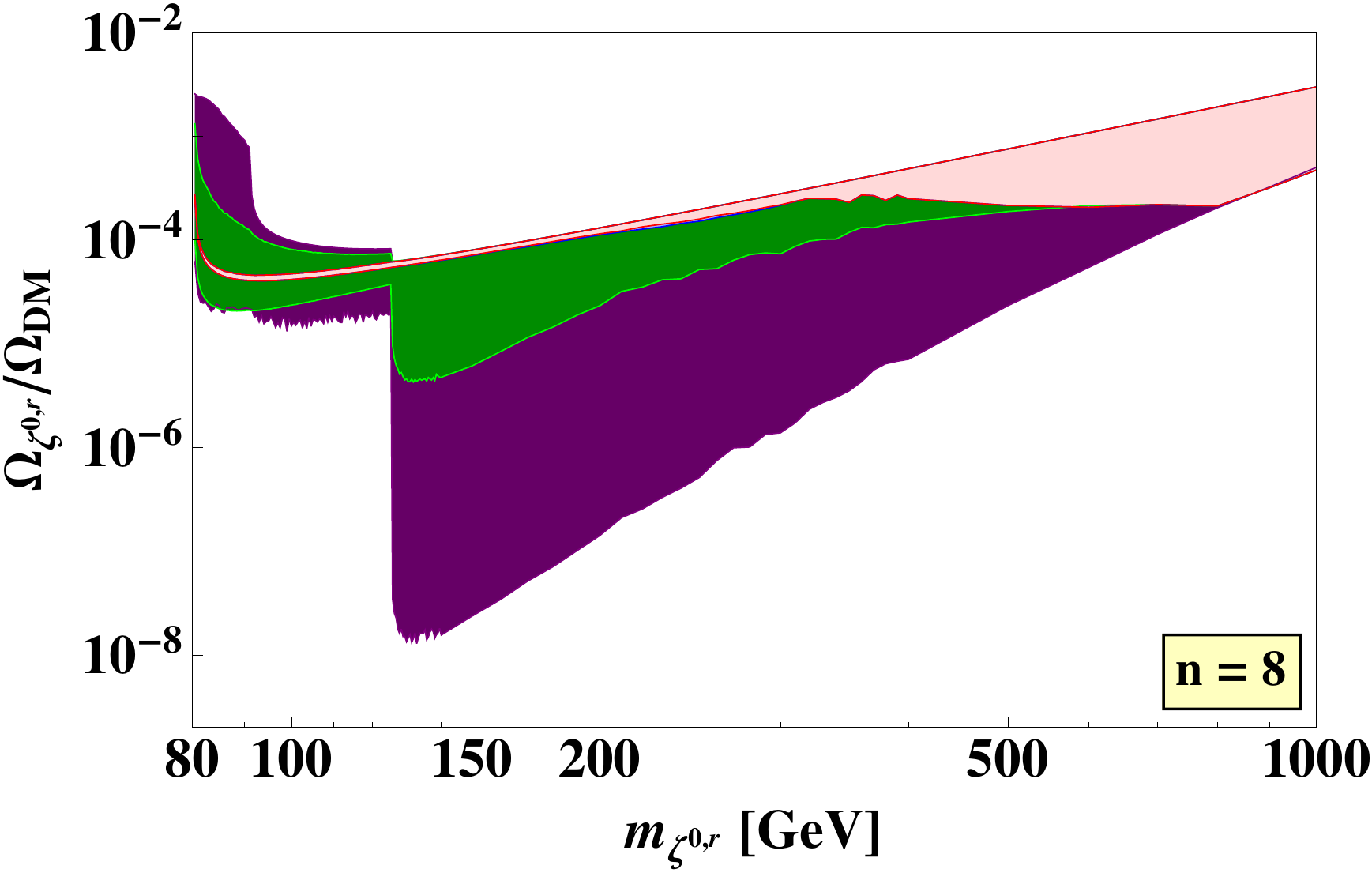}
\caption{The fraction $\Omega_{\zeta^{0,r}}/\Omega_{\rm DM}$ of the total dark matter density that can be made up of $\zeta^{0,r}$ as a function of $m_{\zeta^{0,r}}$.  The colored regions show the accessible range of $\Omega_{\zeta^{0,r}}/\Omega_{\rm DM}$ obtained from a numerical scan over parameter space after successively applying constraints from perturbative unitarity (dark purple), the oblique parameters $S$, $T$, and $U$ (medium green), Higgs decays to two photons (light blue), and the requirement $M^2 > 0$ to avoid alternate minima (very light pink).  Left: $n=6$ model.  Right: $n=8$ model.  (In the $n=8$ case, the very light pink region almost entirely covers the light blue region.)}
\label{fig:omegaplots}
\end{center}
\end{figure}

In Fig.~\ref{fig:omegaplots} we show the fraction of the total dark matter density that can be made up of $\zeta^{0,r}$, computed using Eq.~(\ref{eq:relicratio}) including all kinematically accessible two-body final states, as a function of $m_{\zeta^{0,r}}$ for the models with $n=6$ and 8.  We scan over $\lambda_2$, $\lambda_3$, and $\lambda_4$, applying in succession the constraints from perturbative unitarity (dark purple regions), the oblique parameters $S$, $T$, and $U$ (medium green regions), and Higgs decays to two photons (light blue regions).  We also show the allowed region after imposing all of the above constraints together with the requirement $M^2 > 0$, so that the potential is guaranteed not to have any $Z_2$-breaking minima (very light pink regions).  

The possibility of very small relic densities that opens up when $m_{\zeta^{0,r}} > 125$~GeV is due to the crossing of the kinematic threshold for $\zeta^{0,r} \zeta^{0,r} \to hh$.  However, imposition of the oblique parameter constraints (medium green) and particularly the $h \to \gamma\gamma$ (light blue) and $M^2 > 0$ (very light pink) constraints severely limits the allowed strength $\Lambda_n$ of the $h (h) \zeta^{0,r} \zeta^{0,r}$ couplings, leading to a more tightly constrained relic density.  For $\zeta^{0,r}$ masses above about 500~GeV in the $n=6$ model (800~GeV in the $n=8$ model) the constraints on $\lambda_2$, $\lambda_3$, and $\lambda_4$ come only from perturbative unitarity.

We find that, for allowed parameter choices and $m_{\zeta^{0,r}} \lesssim 1$~TeV, the thermal relic abundance of $\zeta^{0,r}$ can account for at most 1\% of the dark matter.  In particular, the two models that we study are consistent with the observed dark matter relic abundance and thus viable extensions of the SM, assuming that most of the dark matter is made up of some other candidate particle.  

Extending our calculation to higher masses, we find that $\zeta^{0,r}$ could account for all the dark matter for masses of 10--33~TeV for the $n=6$ model, or 18--30~TeV for the $n=8$ model.  However, we have not included effects from co-annihilations or Sommerfeld enhancement which may become important for such heavy masses.  Since we are primarily interested in the viability of the large-multiplet models as candidates for LHC searches below the 1~TeV mass range, a detailed treatment of these issues is beyond the scope of this paper.

\section{Dark matter direct detection}
\label{sec:dirdet}

Scattering of $\zeta^{0,r}$ off a nucleon $N = p, n$ proceeds only via Higgs exchange.  The resulting spin-independent per-nucleon cross section is given by
\begin{equation}
	\sigma_{\rm SI}^{\zeta} = \frac{(f^h_N)^2 \Lambda_n^2 v^2}{4 \pi M_h^4}
	\frac{m_N^2}{(m_N + m_{\zeta^{0,r}})^2},
	\label{eq:ddh}
\end{equation}
where $\Lambda_n^2 v^2$ is the square of the $h\zeta^{0,r}\zeta^{0,r}$ coupling defined in Eq.~(\ref{eq:lambdan}) and the Higgs-nucleon Yukawa couplings are given by~\cite{Ellis:2000ds}
\begin{eqnarray}
	f_p^h &=& \frac{m_p}{v}(0.350 \pm 0.048), \nonumber \\
	f_n^h &=& \frac{m_n}{v}(0.353 \pm 0.049).
\end{eqnarray}
Because the relic density of $\zeta^{0,r}$ is only a fraction of the total dark matter density, the direct-detection scattering cross section $\sigma_{\rm SI}^{\rm exp}$ quoted by experiments does not correspond directly to $\sigma_{\rm SI}^{\zeta}$, but rather to $\sigma_{\rm SI}^{\zeta}$ scaled by the fraction of the total dark matter density that is made up by $\zeta^{0,r}$:
\begin{equation}
	\sigma_{\rm SI}^{\rm exp} 
	= \frac{\Omega_{\zeta^{0,r}}}{\Omega_{\rm DM}} \sigma_{\rm SI}^{\zeta} 
	= \frac{\langle\sigma v_{rel}\rangle_{\rm std}}
	{\langle\sigma v_{rel}(\zeta^{0,r}\zeta^{0,r} \to {\rm any})\rangle}\,\sigma^{\zeta}_{\rm SI},
	\label{eq:sigmaexp}
\end{equation}
where we have used the relic density scaling relationship of Eq.~(\ref{eq:relicratio}).

\begin{figure}
\begin{center}
\includegraphics[width=0.49\textwidth]{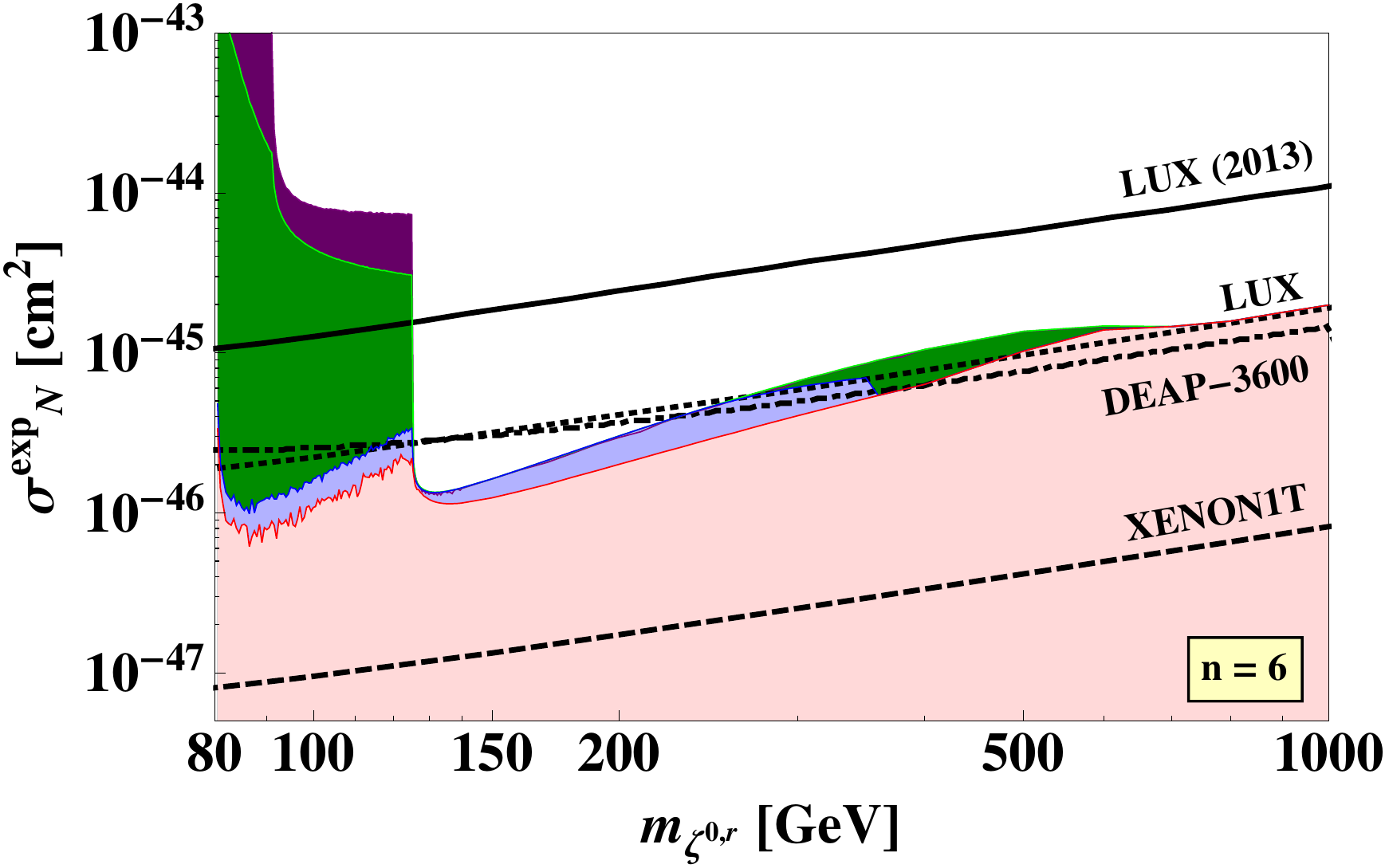}
\includegraphics[width=0.49\textwidth]{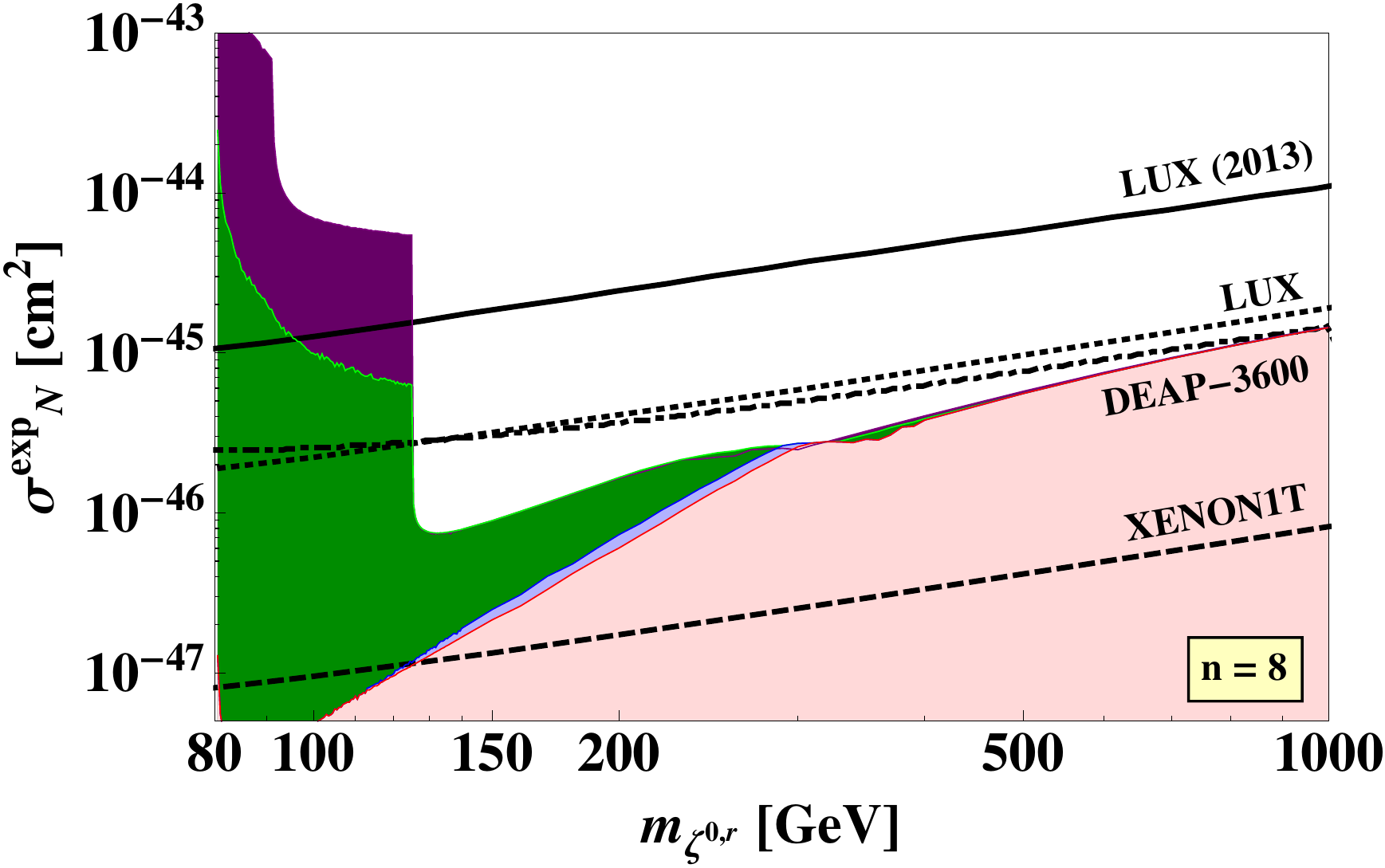}
\caption{Predictions for the experimentally-defined direct detection cross section $\sigma_{\rm SI}^{\rm exp}$ [see Eq.~(\ref{eq:sigmaexp})] as a function of $m_{\zeta^{0,r}}$.  The colored regions show the accessible range of $\sigma_{\rm SI}^{\rm exp}$ obtained from a numerical scan over parameter space after successively applying constraints from perturbative unitarity (dark purple), the oblique parameters $S$, $T$, and $U$ (medium green), Higgs decays to two photons (light blue), and the requirement $M^2 > 0$ to avoid alternate minima (very light pink).  Also shown are the current experimental upper bound from the LUX experiment~\cite{Akerib:2013tjd} (solid black curve), as well as projections for the ultimate sensitivities of the LUX, DEAP-3600, and XENON1T experiments~\cite{DMTools} (dashed curves).  
Left: $n=6$ model.  Right: $n=8$ model.}
\label{fig:sigmaplots}
\end{center}
\end{figure}

In Fig.~\ref{fig:sigmaplots} we show the predicted range of density-scaled experimental direct detection cross sections $\sigma_{\rm SI}^{\rm exp}$, calculated according to Eq.~(\ref{eq:sigmaexp}), as a function of $m_{\zeta^{0,r}}$ for the models with $n=6$ and 8.  As before, we scan over $\lambda_2$, $\lambda_3$, and $\lambda_4$, applying in succession the constraints from perturbative unitarity (dark purple regions), the oblique parameters $S$, $T$, and $U$ (medium green regions), Higgs decays to two photons (light blue regions), and $M^2 > 0$ to avoid alternate minima (very light pink regions).  Because the scattering cross section mediated by Higgs exchange is proportional to $\Lambda_n^2$, $\sigma_{\rm SI}^{\rm exp}$ can be made arbitrarily small by tuning $\lambda_2$, $\lambda_3$ and $\lambda_4$ so that $\Lambda_n \ll 1$.\footnote{In the limit $\Lambda_n \to 0$, loop-induced scattering processes mediated by $W$ and $Z$ bosons will generically contribute at the level of $\sigma_{\rm SI}^{\zeta} \sim 10^{-(46-48)}$~cm$^2$~\cite{Hisano:2011cs}, or $\sigma_{\rm SI}^{\rm exp} \sim 10^{-(48-53)}$~cm$^2$.}

For $\zeta^{0,r}$ masses in our range of interest, the strongest experimental upper limit on $\sigma_{\rm SI}^{\rm exp}$ currently comes from the LUX experiment~\cite{Akerib:2013tjd}.  As can be seen in Fig.~\ref{fig:sigmaplots}, the current sensitivity is not yet sufficient to probe the allowed parameter space after imposing the other constraints on model parameters.  We also show projections for the future sensitivities of the LUX~\cite{Akerib:2012ys}, DEAP-3600~\cite{Boulay:2012hq}, and XENON1T~\cite{Aprile:2012zx} experiments as compiled by the DMTools website~\cite{DMTools}.\footnote{The datasets from DMTools plotted in Fig.~\ref{fig:sigmaplots} are as follows. LUX: 300-day projection, R.~Gaitskell (Brown) and D.~McKinsey (Yale), 2013.  DEAP-3600: 1000~kg fiducial mass, D.~McKinsey (Fermilab), May 2007 projection.  XENON1T: 3~ton-years, K.~Ni (Columbia), 2009 projection.}  The projected LUX and DEAP-3600 sensitivities begin to probe the upper edge of the $n=6$ model.\footnote{We note that more recent compilations such as Ref.~\cite{Cushman:2013zza} quote the DEAP-3600 sensitivity as being a factor of two better than the 2007 projection available from DMTools and shown in Fig.~\ref{fig:sigmaplots}.}  The projected XENON1T sensitivity probes deep into the allowed model parameter space, reaching a full order of magnitude beyond the largest experimental direct-detection cross sections allowed in both the $n=6$ and $n=8$ models.  The projected sensitivity of the SuperCDMS experiment at SNOLAB is close to that of XENON1T~\cite{Cushman:2013zza}.  The projected sensitivity of the proposed LUX upgrade LZ reaches an order of magnitude deeper in $\sigma_{\rm SI}^{\rm exp}$ than that of XENON1T~\cite{Cushman:2013zza}.

Finally, we briefly comment on inelastic direct-detection scattering processes.  In the high-mass region, where the mass splittings are small, it may be possible for the $\zeta^{0,r}$ to up-scatter to a $\zeta^{0,i}$, providing an alternative direct detection mechanism through $Z$-boson exchange. At the kinematic threshold, the accessible mass splitting is given in terms of the ambient $\zeta^{0,r}$ velocity $v_{\zeta}$ by
\begin{equation}
	m_{\zeta^{0,i}} - m_{\zeta^{0,r}} 
	\leq \sqrt{m_A^2 + m_{\zeta^{0,r}}^2 + 2m_A m_{\zeta^{0,r}}\sqrt{1+v_{\zeta}^2}} 
	- (m_A + m_{\zeta^{0,r}}),
	\label{eq:upscatter}
\end{equation}
where $m_A$ is the mass of the scattering target. In our case, the current dark matter velocity is $v_{\zeta} \sim 10^{-3}c$ and the scattering target is a xenon nucleus with $m_A \approx 124$~GeV.  For $m_{\zeta^{0,r}} \gg m_A$, Eq.~(\ref{eq:upscatter}) approaches a constant value, $m_{\zeta^{0,i}} - m_{\zeta^{0,r}} \lesssim m_A v_{\zeta}^2/2 \sim 60$~keV.  In turn, this implies that the mass splittings between the electrically-charged scalars and $\zeta^{0,r}$ would be much less than the mass of an electron, forcing the electrically-charged scalars to be stable.  We conclude that the absence of heavy charged relics precludes the possibility of inelastic direct-detection scattering in our models.

\section{Discussion and conclusions}
\label{sec:conclusions}

In this paper we examined extensions of the SM scalar sector containing a single large multiplet of SU(2)$_L$ in addition to the SM Higgs doublet.  We focused on two models, one in which the large multiplet has isospin $T = 5/2$ ($n=6$) and hypercharge $Y=1$ and the other with $T=7/2$ ($n=8$) and $Y=1$.  We impose a global $Z_2$ symmetry on the scalar potential, which forces the lightest member of the large multiplet to be stable.

Starting from the scalar potential for each of the two models, we worked out the spectrum of mass eigenstates and their couplings to SM gauge and Higgs bosons.  We then determined the constraints on the model parameters from perturbative unitarity, the oblique parameters $S$, $T$, and $U$, and SM Higgs decays to two photons.  We also imposed a sufficient condition on the $Z$ mass-squared parameter to ensure that the potential has no $Z_2$-breaking minima.  We computed the predictions for Higgs decays to $Z \gamma$, as well as the thermal relic abundance of the lightest member of the large multiplet and its cross section in dark matter direct-detection experiments.

We found that both models are viable for a wide range of masses of the new scalars within the kinematic reach of the LHC.  The mass splittings of the new scalars are constrained mainly by the oblique parameters, which force the mass eigenstates to be tightly clustered into two groups.  These groups can in turn be separated by tens to hundreds of GeV, depending on the mass of the lightest new scalar.  This feature of the spectrum will have interesting implications for the kinematics of collider events involving pair production and decay of the new scalars.

We finish by commenting on a few features of the models that may warrant further study.
\begin{itemize}

\item The masses of the new scalars are bounded from above by the requirement that their relic density not be larger than the observed dark matter density in the universe.  This bound will lie in the several-to-tens of TeV range.  An accurate determination of this bound will require a more careful treatment of co-annihilations and Sommerfeld effects in the calculation of the relic density. 

\item The total cross section for production of pairs of the new scalars at the LHC via electroweak processes will be enhanced by the large multiplicity of scalar states and by their large weak charges.  The LHC reach for these particles may thus extend to higher masses than the reach for other scalar extensions of the SM involving smaller representations of SU(2)$_L$.

\item Our models affect the running of the electroweak gauge couplings.  In particular, the one-loop SU(2)$_L$ beta function coefficient becomes $b_2 = -19/6 + n(n^2-1)/36$, where $\alpha_2^{-1}(\Lambda) = \alpha_2^{-1}(M_Z) - (b_2/2\pi)\log(\Lambda/M_Z)$ and $\alpha_2 \equiv g^2/4\pi$.  In the $n=6$ model, $\alpha_2$ remains perturbative up to well beyond the Planck scale.  In the $n=8$ model, $\alpha_2$ becomes nonperturbative around $10^{10}$~GeV.  The $n=8$ model can be saved by, e.g., lowering the Planck scale below $10^{10}$~GeV through the introduction of flat or warped extra dimensions, or by making the scalars in the large multiplet be composites of fermions which individually transform under smaller representations of SU(2)$_L$.  

\item Our models suffer from the hierarchy problem in the same way as does the SM Higgs sector.  The $n=8$ model cannot be supersymmetrized because the addition of a second $n=8$ multiplet with $Y=-1$, as required for anomaly cancellation, would violate perturbative unitarity in transversely-polarized $WW \to \zeta\zeta$ scattering amplitudes~\cite{Hally:2012pu}.  Supersymmetrizing the $n=6$ model modifies the one-loop SU(2)$_L$ beta function coefficient to read $b_2 = 1 + 6 \cdot n(n^2-1)/36$; for supersymmetry at the weak scale, $\alpha_2$ becomes nonperturbative around $10^5$~GeV.  The hierarchy problem could be solved in either model through compositeness of the scalar fields.

\item The presence of the large multiplet coupled to the SM Higgs doublet affects the running of the SM Higgs quartic coupling, here identified with $\lambda_1$.  Further constraints on the parameter space could be imposed by requiring that the quartic couplings remain perturbative and the vacuum remains (meta)stable up to a chosen cutoff scale.  We note, however, that any such constraints would be subject to the assumption that no new physics---including physics needed to solve the hierarchy problem---enters below the cutoff scale.

\item We have assumed that the quartic couplings in the $\mathcal{O}(Z^4)$ part of the scalar potential in Eq.~(\ref{eq:conjugatepotential}) can always be chosen so that the potential is bounded from below.  The sizes of these quartic couplings will, however, be constrained by perturbative unitarity.  The interplay of this constraint with the requirement that the potential be bounded from below may further constrain the allowed ranges of $\lambda_2$, $\lambda_3$, and $\lambda_4$.

\end{itemize}

\begin{acknowledgments}
We thank Thomas Gr\'egoire, Pat Kalyniak, Travis Martin, Guy Moore, and Brooks Thomas for enlightening discussions.
This work was supported by the Natural Sciences and Engineering Research Council of
Canada.  K.H.\ was also supported by the Government of Ontario through an Ontario Graduate Scholarship.
\end{acknowledgments}

\appendix

\section{Masses and mixing angles}
\label{app:models}

In this section we give some of the mathematical details used in the derivation of the mass spectrum and mixing angles in Sec.~\ref{sec:models}.

For a complex scalar multiplet $Z$ with hypercharge $Y=1$ (normalized so that $Q = T^3 + Y/2$), the most general gauge-invariant and $Z_2$-invariant renormalizable scalar potential was given in Eq.~(\ref{eq:conjugatepotential}), in which $\widetilde \Phi =  i \sigma^2 \Phi^*$ and $\widetilde Z = C Z^*$ are the conjugate multiplets.  Here $\sigma^2$ is the second Pauli matrix and
the conjugation matrix $C$ for the large multiplet is an anti-diagonal $n \times n$ matrix. For $n=6$ and 8 it is given by
\begin{equation}
C_{(n = 6)} = \left(\begin{matrix} 0 & 0 & 0 & 0 & 0 & 1\\
0 & 0 & 0 & 0 & -1 & 0\\
0 & 0 & 0 & 1 & 0 & 0\\
0 & 0 & -1 & 0 & 0 & 0\\
0 & 1 & 0 & 0 & 0 & 0\\
-1 & 0 & 0 & 0 & 0 & 0
\end{matrix}\right),  \qquad
C_{(n = 8)} = \left(\begin{matrix}0 & 0 & 0 & 0 & 0 & 0 & 0 & 1\\
0 & 0 & 0 & 0 & 0 & 0 & -1 & 0\\
0 & 0 & 0 & 0 & 0 & 1 & 0 & 0\\
0 & 0 & 0 & 0 & -1 & 0 & 0 & 0\\
0 & 0 & 0 & 1 & 0 & 0 & 0 & 0\\
0 & 0 & -1 & 0 & 0 & 0 & 0 & 0\\
0 & 1 & 0 & 0 & 0 & 0 & 0 & 0\\
-1 & 0 & 0 & 0 & 0 & 0 & 0 & 0
\end{matrix}\right).
\label{eq:Cmatrix}
\end{equation}

Taking $\lambda_4$ real and working in unitarity gauge, the term involving $\lambda_4$ in the scalar potential of Eq.~(\ref{eq:conjugatepotential}) reduces to
\begin{equation}
	\lambda_4\, \widetilde{\Phi}^\dag \tau^a \Phi\, Z^\dag T^a \widetilde{Z} + \mathrm{h.c.} 
	= \frac{1}{4}\lambda_4 (h + v)^2\left[Z^\dag T^{-} \widetilde{Z} 
	+ \widetilde{Z}^\dag T^{+} Z\right],
\end{equation}
where $T^{\pm} = T^1 \pm i T^2$.  The terms $Z^\dag T^{-} \widetilde{Z}$ and $\widetilde{Z}^\dag T^{+} Z$ split the masses of $\zeta^{0,r}$ and $\zeta^{0,i}$ and cause mixing between states with the same electric charge but different isospin.
For $n=6,8$, the two pieces can be written as
\begin{eqnarray}
	Z^\dag T^{-} \widetilde{Z} &=& \frac{n}{2}(-1)^{n/2+1}\zeta^{0*}\zeta^{0*} + \sum_{Q=1}^{n-1} \sqrt{n^2-4Q^2} (-1)^{n/2+Q+1}\,\zeta^{+Q*}\zeta^{-Q*}, \nonumber \\
	\widetilde{Z}^\dag T^{+} Z &=& \frac{n}{2}(-1)^{n/2+1}\zeta^{0}\zeta^{0} + \sum_{Q=1}^{n-1} \sqrt{n^2-4Q^2} (-1)^{n/2+Q+1}\,\zeta^{+Q}\zeta^{-Q}.
\end{eqnarray}
When the neutral state $\zeta^0$ is written in terms of its real and imaginary components, $\zeta^0 = (\zeta^{0,r} + i\zeta^{0,i})/\sqrt{2}$, we find a mass splitting between the components,
\begin{eqnarray}
	m_{\zeta^{0,r}}^2 &=& M^2 + \frac{1}{2}v^2\left[\lambda_2 + \frac{1}{4}\lambda_3 + \frac{n}{2}(-1)^{n/2+1}\lambda_4\right] \equiv M^2 + \frac{1}{2}v^2\Lambda_n, \nonumber \\
	m_{\zeta^{0,i}}^2 &=& M^2 + \frac{1}{2}v^2\left[\lambda_2 + \frac{1}{4}\lambda_3 + \frac{n}{2}(-1)^{n/2}\lambda_4\right] = m_{\zeta^{0,r}}^2 + \frac{n}{2}(-1)^{n/2}v^2\lambda_4.
\end{eqnarray}

The mass matrices for the pairs of scalars with electric charge $Q=1, \ldots, n/2-1$ are given in the basis $(\zeta^{+Q}, \zeta^{-Q*})$ by
\begin{equation}
	M_{\zeta^{\pm Q}}^2 
	= \left(\begin{matrix}M^2 + \frac{1}{8}v^2(4\lambda_2 - (2Q-1)\lambda_3) & \frac{1}{4}v^2\lambda_4\sqrt{n^2-4Q^2}\,(-1)^{n/2+Q+1}\\ \frac{1}{4} v^2\lambda_4\sqrt{n^2-4Q^2}\,(-1)^{n/2+Q+1}  & M^2 + \frac{1}{8}v^2(4\lambda_2 + (2Q+1)\lambda_3)\end{matrix}\right),
\end{equation}
which we diagonalize to find the mass eigenvalues,
\begin{eqnarray}
	m_{H_{1,2}^{Q}}^2 &=& M^2 + \frac{1}{2}v^2\left(\lambda_2 + \frac{1}{4}\lambda_3 \mp \frac{1}{2}\sqrt{Q^2\lambda_3^2 + (n^2-4Q^2)\lambda_4^2}\right) \nonumber \\
	&=& m_{\zeta^{0,r}}^2 + \frac{1}{4}v^2\left(n(-1)^{n/2}\lambda_4 \mp \sqrt{Q^2\lambda_3^2 + (n^2-4Q^2)\lambda_4^2}\right).
\end{eqnarray}
The mass eigenstates $H_1^{Q}$ and $H_2^{Q}$ are defined in terms of the weak eigenstates by Eq.~(\ref{eq:alphadef}) such that $H_1^{+Q}$ is the lighter state and $H_2^{+Q}$ is the heavier state.
The mixing angle $\alpha_Q \in\nobreak [-\frac{\pi}{2},\frac{\pi}{2}]$ is given by
\begin{eqnarray}
	\tan \alpha_Q &=& (-1)^{n/2+Q+1}\frac{Q\lambda_3 - \sqrt{Q^2\lambda_3^2 + (n^2-4Q^2)\lambda_4^2}}{\sqrt{n^2-4Q^2}\lambda_4} \nonumber \\
	&=& (-1)^{n/2+Q}\frac{\sqrt{n^2-4Q^2}\lambda_4}{Q\lambda_3 + \sqrt{Q^2\lambda_3^2 + (n^2-4Q^2) \lambda_4^2}}.
\end{eqnarray}

There is only one state with $Q=n/2$.  Its mass is given by
\begin{equation}
	m_{\zeta^{n/2}}^2 = M^2 + \frac{1}{8}v^2\left(4\lambda_2 - (2Q-1)\lambda_3\right) 
	= m_{\zeta^{0,r}}^2 - \frac{n}{8}v^2\left(\lambda_3 + 2 (-1)^{n/2+1}\lambda_4\right).
\end{equation}

\section{Feynman rules}
\label{sec:rulesB}

In this section we collect the Feynman rules for the couplings of the new scalars to gauge and Higgs bosons.  We define the couplings with all particles and momenta incoming.  For couplings involving scalar momenta, we define $p_1$ as the momentum of the first scalar and $p_2$ as the momentum of the second scalar.  

For simplicity in the derivation of the oblique parameters, all coefficients $C$ for couplings of scalars to one or two electroweak gauge bosons are defined with the overall factors of $e$ removed: one factor of $e$ is removed from couplings to a single gauge boson and two factors of $e$ are removed from couplings to two gauge bosons.

\subsection{Higgs boson couplings to scalar pairs}
\label{sec:FRhiggs}

The Feynman rule for the coupling $h s_1 s_2$ is given by $-i C_{h s_1s_2}$, where
\begin{eqnarray}
	C_{h \zeta^{0,r} \zeta^{0,r}} &=& v \left(\lambda_2 + \frac{1}{4}\lambda_3 + \frac{n}{2} (-1)^{n/2+1} \lambda_4\right), \nonumber \\
	C_{h \zeta^{0,i} \zeta^{0,i}} &=& v\left(\lambda_2 + \frac{1}{4}\lambda_3 + \frac{n}{2} (-1)^{n/2}\,\lambda_4\right), \nonumber \\
	C_{h H_1^Q H_1^{-Q}} &=& v\left(\lambda_2 + \frac{1}{4}\lambda_3 - \frac{1}{2}\sqrt{Q^2\lambda_3^2 + (n^2-4Q^2)\lambda_4^2}\right), \nonumber \\ 
	C_{h H_2^Q H_2^{-Q}} &=& v\left(\lambda_2 + \frac{1}{4}\lambda_3 + \frac{1}{2}\sqrt{Q^2\lambda_3^2 + (n^2-4Q^2)\lambda_4^2}\right), \nonumber \\ 
	C_{h \zeta^{n/2} \zeta^{-n/2}} &=& v\left(\lambda_2 - \frac{2Q-1}{4}\lambda_3\right).
	\label{eq:hH1H1} 
\end{eqnarray}
The Feynman rule for the coupling $hh s_1 s_2$ is given by $-i C_{hh s_1s_2}$, where 
\begin{eqnarray}
	C_{hh \zeta^{0,r} \zeta^{0,r}} &=& \lambda_2 + \frac{1}{4}\lambda_3 + \frac{n}{2} (-1)^{n/2+1}\,\lambda_4, \nonumber \\
	C_{hh \zeta^{0,i} \zeta^{0,i}} &=& \lambda_2 + \frac{1}{4}\lambda_3 + \frac{n}{2} (-1)^{n/2}\,\lambda_4, \nonumber \\
	C_{hh H_1^Q H_1^{-Q}} &=& \lambda_2 + \frac{1}{4}\lambda_3 - \frac{1}{2}\sqrt{Q^2\lambda_3^2 + (n^2-4Q^2)\lambda_4^2}, \nonumber \\
	C_{hh H_2^Q H_2^{-Q}} &=& \lambda_2 + \frac{1}{4}\lambda_3 + \frac{1}{2}\sqrt{Q^2\lambda_3^2 + (n^2-4Q^2)\lambda_4^2}, \nonumber \\
	C_{hh \zeta^{n/2} \zeta^{-n/2}} &=& \lambda_2 - \frac{2Q-1}{4}\lambda_3.
	\label{eq:hhH1H1}
\end{eqnarray}
Note that, for all the couplings above, $s_2 = s_1^*$; i.e., there are no off-diagonal couplings.

\subsection{Gauge boson couplings to scalar pairs}
\label{sec:FRgauge}

The Feynman rules for the couplings of the new scalars to gauge bosons come from the gauge-kinetic terms in the Lagrangian,
\begin{equation}
	\mathcal{L} \supset \left( \mathcal{D}_{\mu} Z \right)^{\dagger} \left( \mathcal{D}^{\mu} Z \right),
\end{equation}
where the covariant derivative is given by
\begin{equation}
	\mathcal{D}_{\mu} = \partial_{\mu} - i \frac{g}{\sqrt{2}} \left( W_{\mu}^+ T^+ + W_{\mu}^- T^- \right) - i \frac{e}{s_W c_W} Z_{\mu} \left( T^3 - s_W^2 Q \right) - i e A_{\mu} Q.
\end{equation}

\subsubsection{Couplings to one or two photons}

The Feynman rule for the coupling $s_1 s_2 \gamma_{\mu}$, for $s_1$ with charge $Q$ and $s_2 = s_1^*$, is
\begin{equation}
	ie C_{s_1 s_2 \gamma}(p_1 - p_2)_{\mu}, \qquad {\rm where} \quad  C_{s_1 s_2 \gamma} = Q.
\end{equation}
The Feynman rule for the coupling $s_1 s_2 \gamma_{\mu} \gamma_{\nu}$, for $s_1$ with charge $Q$ and $s_2 = s_1^*$, is	
\begin{equation}
	-i e^2 C_{s_1 s_2 \gamma\gamma} g_{\mu\nu}, \qquad {\rm where} \quad C_{s_1 s_2 \gamma\gamma} = -2 Q^2.
\end{equation}
There are no off-diagonal couplings, in accordance with the conservation of the electromagnetic current.

\subsubsection{Couplings to one $Z$ boson}

The Feynman rule for the coupling $s_1 s_2 Z_{\mu}$ is given by
\begin{equation}
	i e C_{s_1 s_2 Z} (p_1 - p_2)_{\mu},
\end{equation}
where
\begin{eqnarray}
	C_{\zeta^{0,r} \zeta^{0,i} Z} &=& \frac{i}{2 s_Wc_W}, \nonumber \\
	C_{H_1^Q H_1^{-Q} Z} &=& \frac{1}{s_Wc_W} \left[ \left(Q-\frac{1}{2}\right) \cos^2\alpha_Q + \left(Q+\frac{1}{2}\right) \sin^2 \alpha_Q - Q s_W^2 \right], \nonumber \\
	C_{H_2^Q H_2^{-Q} Z} &=& \frac{1}{s_Wc_W} \left[ \left(Q-\frac{1}{2}\right) \sin^2\alpha_Q + \left(Q+\frac{1}{2}\right) \cos^2 \alpha_Q - Q s_W^2 \right], \nonumber \\
	C_{H_1^Q H_2^{-Q} Z} = C_{H_2^Q H_1^{-Q} Z} &=& \frac{1}{s_W c_W}\sin{\alpha_Q}\cos{\alpha_Q}, \nonumber \\
	C_{\zeta^{n/2} \zeta^{-n/2} Z} &=& \frac{1}{s_W c_W}\left[\frac{n-1}{2}-\frac{n}{2}s_W^2\right].
\end{eqnarray}
Note that the diagonal couplings $C_{\zeta^{0,r} \zeta^{0,r} Z} = C_{\zeta^{0,i} \zeta^{0,i} Z} = 0$ due to parity conservation.

\subsubsection{Couplings to ZZ}

The Feynman rule for the coupling $s_1 s_2 Z_{\mu} Z_{\nu}$ is given by
\begin{equation}
	-i e^2 C_{s_1 s_2 ZZ} g_{\mu\nu},
\end{equation}
where
\begin{eqnarray}
	C_{\zeta^{0,r} \zeta^{0,r} ZZ} = C_{\zeta^{0,i} \zeta^{0,i} ZZ} &=& -\frac{1}{2 s_W^2c_W^2}, \nonumber \\
	C_{H_1^Q H_1^{-Q} ZZ} &=& -\frac{2}{s_W^2 c_W^2} \left[ \left(Q c_W^2-\frac{1}{2}\right)^2 \cos^2\alpha_Q + \left(Q c_W^2+\frac{1}{2}\right)^2 \sin^2\alpha_Q \right], \nonumber \\
	C_{H_1^Q H_2^{-Q} ZZ} = C_{H_2^Q H_1^{-Q} ZZ} &=& -\frac{4 Q}{s_W^2 c_W^2} (1 - s_W^2) \sin\alpha_Q \cos\alpha_Q, \nonumber \\
	C_{H_2^Q H_2^{-Q} ZZ} &=& -\frac{2}{s_W^2 c_W^2} \left[ \left(Q c_W^2-\frac{1}{2}\right)^2 \sin^2\alpha_Q + \left(Q c_W^2+\frac{1}{2}\right)^2 \cos^2\alpha_Q \right], \nonumber \\
	C_{\zeta^{n/2} \zeta^{-n/2} ZZ} &=& -\frac{2}{s_W^2c_W^2} \left[\frac{n-1}{2} - \frac{n}{2} s_W^2 \right]^2.
\end{eqnarray}
Note that the off-diagonal coupling $\zeta^{0,r} \zeta^{0,i} ZZ$ is zero.

\subsubsection{Couplings to Z$\gamma$}

The Feynman rule for the coupling $s_1 s_2 Z_{\mu} \gamma_{\nu}$ is given by
\begin{equation}
	-i e^2 C_{s_1 s_2 Z \gamma} g_{\mu\nu},
\end{equation}
where
\begin{eqnarray}
	C_{H_1^Q H_1^{-Q} Z\gamma} &=& -\frac{2 Q}{s_Wc_W} \left[\left(Q-\frac{1}{2}\right) \cos^2\alpha_Q + \left(Q+\frac{1}{2}\right) \sin^2 \alpha_Q - Q s_W^2 \right], \nonumber \\
	C_{H_1^Q H_2^{-Q} Z\gamma} = C_{H_2^Q H_1^{-Q} Z\gamma} &=& -\frac{2 Q}{s_W c_W} \sin\alpha_Q \cos\alpha_Q, \nonumber \\
	C_{H_2^Q H_2^{-Q} Z\gamma} &=& -\frac{2 Q}{s_Wc_W} \left[\left(Q-\frac{1}{2}\right) \sin^2\alpha_Q + \left(Q+\frac{1}{2}\right)+\cos^2 \alpha_Q - Q s_W^2 \right], \nonumber \\
	C_{\zeta^{n/2} \zeta^{-n/2} Z\gamma} &=& -\frac{n e^2}{s_Wc_W} \left[ \frac{n-1}{2} - \frac{n}{2} s_W^2 \right].
\end{eqnarray}
The neutral scalars do not couple to $Z\gamma$.

\subsubsection{Couplings to one W boson}

The Feynman rule for the coupling $s_1 s_2 W^{\pm}_{\mu}$ is given by
\begin{equation}
	i e C_{s_1 s_2 W^{\pm}} (p_1 - p_2)_{\mu}.
\end{equation}
For compactness, we define the following coefficients for a given value of $n$:
\begin{eqnarray}
	T^+_Q &=& \frac{1}{2}\sqrt{n^2 - 4 Q^2}, \nonumber \\
	T^-_Q &=& \frac{1}{2}\sqrt{n^2 - 4 (Q-1)^2}. 
	\label{Tpm}
\end{eqnarray}
Then the couplings of two scalars to $W^+$ are given by 
\begin{eqnarray}
	C_{\zeta^{0,r} H_1^- W^+} &=& \frac{1}{2 s_W}\left[ \frac{n}{2} \cos\alpha_1 - T_{-1}^+ \sin\alpha_1 \right], \nonumber \\
	C_{\zeta^{0,r} H_2^- W^+} &=& \frac{1}{2 s_W}\left[ -\frac{n}{2} \sin\alpha_1 - T_{-1}^+ \cos\alpha_1 \right], \nonumber \\
	C_{\zeta^{0,i} H_1^- W^+} &=& \frac{i}{2 s_W} \left[ \frac{n}{2} \cos\alpha_1 + T_{-1}^+ \sin\alpha_1 \right], \nonumber \\
	C_{\zeta^{0,i} H_2^- W^+} &=& \frac{i}{2 s_W} \left[ -\frac{n}{2} \sin\alpha_1 + T_{-1}^+ \cos\alpha_1 \right], \nonumber \\
	C_{H_1^Q H_1^{-(Q+1)} W^+} &=& \frac{1}{\sqrt{2} s_W} \left[ T_Q^+ \cos\alpha_Q \cos\alpha_{Q+1} - T_{-Q-1}^+ \sin\alpha_Q \sin\alpha_{Q+1} \right], \nonumber \\
	C_{H_1^Q H_2^{-(Q+1)} W^+} &=& \frac{1}{\sqrt{2} s_W}  \left[ -T_Q^+ \cos\alpha_Q \sin\alpha_{Q+1} - T_{-Q-1}^+ \sin\alpha_Q \cos\alpha_{Q+1} \right], \nonumber \\
	C_{H_2^Q H_1^{-(Q+1)} W^+} &=& \frac{1}{\sqrt{2} s_W}  \left[ -T_Q^+ \sin\alpha_Q \cos\alpha_{Q+1} - T_{-Q-1}^+ \cos\alpha_Q \sin\alpha_{Q+1} \right], \nonumber \\
	C_{H_2^Q H_2^{-(Q+1)} W^+} &=& \frac{1}{\sqrt{2} s_W}  \left[ T_Q^+ \sin\alpha_Q \sin\alpha_{Q+1} - T_{-Q-1}^+ \cos\alpha_Q \cos\alpha_{Q+1} \right], \nonumber \\
	C_{H_1^{n/2-1} \zeta^{-n/2} W^+} &=& \frac{1}{\sqrt{2} s_W} T^+_{n/2-1} \cos\alpha_{n/2-1}, \nonumber \\
	C_{H_2^{n/2-1} \zeta^{-n/2} W^+} &=& -\frac{1}{\sqrt{2} s_W} T^+_{n/2-1} \sin\alpha_{n/2-1}.
\end{eqnarray}
The couplings of two scalars to $W^-$ are obtained using the relation
\begin{equation}
	C_{s_2^* s_1^* W^-} = (C_{s_1 s_2 W^+})^*.
\end{equation}
Note that all the couplings $C_{s_1 s_2 W^+}$ are real except for those that involve one $\zeta^{0,i}$, which are imaginary.

\subsubsection{Couplings to $W^+W^-$}

The Feynman rule for the coupling $s_1 s_2 W^+_{\mu} W^-_{\nu}$ is given by
\begin{equation}
	-i e^2 C_{s_1s_2 W^+W^-} g_{\mu\nu}.
\end{equation}
For compactness we further define, for a given value of $n$, 
\begin{eqnarray}
	T^{+-}_Q &=& T^+_{Q}\,T^-_{Q+1}+T^-_{Q}\,T^+_{Q-1} =\frac{n^2-2}{2}-2Q(Q-1), \nonumber \\
	T^{-+}_Q &=& T^-_{Q}\,T^+_{Q-1}+T^+_{Q}\,T^-_{Q+1} =\frac{n^2-2}{2}-2Q(Q+1).
\end{eqnarray}
Then the couplings of two scalars to $W^+ W^-$ are given by
\begin{eqnarray}
	C_{\zeta^{0,r} \zeta^{0,r} W^+W^-} = C_{\zeta^{0,i} \zeta^{0,i} W^+W^-} &=& - \frac{1}{2 s_W^2} T^{+-}_{0}, \nonumber \\
	C_{H_1^Q H_1^{-Q} W^+W^-} &=& - \frac{1}{2 s_W^2} \left[ T^{+-}_{Q} \cos^2\alpha_Q + T^{-+}_{Q} \sin^2 \alpha_Q \right], \nonumber \\
	C_{H_2^Q H_2^{-Q} W^+W^-} &=& - \frac{1}{2 s_W^2} \left[ T^{+-}_{Q} \sin^2\alpha_2 +T^{-+}_{Q} \cos^2 \alpha_2 \right], \nonumber \\
	C_{H_1^Q H_2^{-Q} W^+W^-} = C_{H_2^Q H_1^{-Q} W^+ W^-} &=& \frac{2 Q}{s_W^2}\sin\alpha_Q \cos\alpha_Q, \nonumber \\
	C_{\zeta^{n/2} \zeta^{-n/2} W^+W^-} &=& - \frac{1}{2 s_W^2} T^{+-}_{n/2}.
\end{eqnarray}
Note that the off-diagonal coupling $\zeta^{0,r} \zeta^{0,i} W^+_{\mu} W^-_{\nu}$ is zero.

\subsubsection{Couplings to $W^+W^+$ and $W^- W^-$}

The Feynman rule for the coupling of two scalars to two like-sign $W$ bosons, $s_1 s_2 W^+_{\mu} W^+_{\nu}$, is given by
\begin{equation}
	-i e^2 C_{s_1s_2W^+W^+} g_{\mu\nu},
\end{equation}
where, for $Q > 0$,
\begin{eqnarray}
	C_{\zeta^{0,r} H_1^{--} W^+W^+} &=& -\frac{1}{\sqrt{2}s_W^2}\left[ T^+_{0} T^-_{2} \cos\alpha_2 + T^+_{-2} T^-_{0} \sin\alpha_2 \right], \nonumber \\
	C_{\zeta^{0,r} H_2^{--} W^+W^+} &=& -\frac{1}{\sqrt{2}s_W^2} \left[ -T^+_{0} T^-_{2} \sin\alpha_2 + T^+_{-2} T^-_{0} \cos\alpha_2 \right], \nonumber \\
	C_{\zeta^{0,i} H_1^{--} W^+W^+} &=& -\frac{i}{\sqrt{2}s_W^2}\left[ T^+_{0} T^-_{2} \cos\alpha_2 - T^+_{-2} T^-_{0}\sin\alpha_2 \right], \nonumber \\
		C_{\zeta^{0,i} H_2^{--} W^+W^+} &=& -\frac{i}{\sqrt{2}s_W^2}\left[ -T^+_{0} T^-_{2}\sin\alpha_2 - T^+_{-2} T^-_{0} \cos\alpha_2 \right], \nonumber \\
	C_{H_1^Q H_1^{-(Q+2)}W^+W^+} &=& -\frac{1}{s_W^2} \left[ T^+_{Q} T^-_{Q+2} \cos\alpha_Q \cos\alpha_{Q+2} + T^+_{-Q-2} T^-_{-Q} \sin\alpha_Q \sin\alpha_{Q+2} \right], \nonumber \\
	C_{H_2^Q H_2^{-(Q+2)} W^+W^+} &=& -\frac{1}{s_W^2}\left[ T^+_{Q} T^-_{Q+2} \sin\alpha_Q \sin\alpha_{Q+2} + T^+_{-Q-2} T^-_{-Q} \cos\alpha_Q \cos\alpha_{Q+2} \right], \nonumber \\
	C_{H_1^Q H_2^{-(Q+2)} W^+W^+} &=& -\frac{1}{s_W^2}\left[ -T^+_{Q} T^-_{Q+2} \cos\alpha_Q \sin\alpha_{Q+2} +T^+_{-Q-2} T^-_{-Q} \sin\alpha_Q \cos\alpha_{Q+2} \right], \nonumber \\
	C_{H_2^Q H_1^{-(Q+2)} W^+W^+} &=& -\frac{1}{s_W^2} \left[ -T^+_{Q} T^-_{Q+2} \sin\alpha_Q \cos\alpha_{Q+2} + T^+_{-Q-2} T^-_{-Q} \cos\alpha_Q \sin\alpha_{Q+2} \right], \nonumber \\
	C_{H_1^{n/2-2} \zeta^{-n/2} W^+W^+} &=& -\frac{1}{s_W^2} T^+_{n/2-2} T^-_{n/2}\cos\alpha_{n/2-2}, \nonumber \\
	C_{H_2^{n/2-2} \zeta^{-n/2} W^+W^+} &=& \frac{1}{s_W^2} T^+_{n/2-2} T^-_{n/2} \sin\alpha_{n/2-2}, \nonumber \\
	C_{H_1^- H_1^- W^+W^+} &=& -\frac{2}{s_W^2} T^+_{1} T^-_{1} \cos\alpha_1 \sin\alpha_1, \nonumber \\
	C_{H_2^- H_2^- W^+W^+} &=& \frac{2}{s_W^2} T^+_{1} T^-_{1} \cos\alpha_1 \sin\alpha_1, \nonumber \\
	C_{H_1^- H_2^- W^+W^+} &=& -\frac{1}{s_W^2}T^+_{1} T^-_{1} ( \cos^2\alpha_1 - \sin^2\alpha_1).
\end{eqnarray}
The couplings of two scalars to $W^-W^-$ are obtained using the relation
\begin{equation}
	C_{s_2^* s_1^* W^- W^-} = (C_{s_1 s_2 W^+ W^+})^*.
\end{equation}
Note that all the couplings $C_{s_1 s_2 W^+W^+}$ are real except for those that involve one $\zeta^{0,i}$, which are imaginary.

\subsubsection{Couplings to $W \gamma$}

The Feynman rule for the coupling $s_1 s_2 W^{\pm}_{\mu} \gamma_{\nu}$ is given by
\begin{equation}
	-i e^2 C_{s_1 s_2 W^{\pm} \gamma} g_{\mu\nu},
\end{equation}
where
\begin{eqnarray}
	C_{\zeta^{0,r} H_1^- W^+\gamma} &=& -\frac{1}{2 s_W}\left[ \frac{n}{2} \cos\alpha_1 - T_{-1}^+ \sin\alpha_1 \right], \nonumber \\
	C_{\zeta^{0,r} H_2^- W^+\gamma} &=& -\frac{1}{2 s_W} \left[ -\frac{n}{2} \sin\alpha_1 - T_{-1}^+ \cos\alpha_1 \right], \nonumber \\
	C_{\zeta^{0,i} H_1^- W^+\gamma} &=& -\frac{i}{2 s_W}\left[ \frac{n}{2} \cos\alpha_1 + T_{-1}^+ \sin\alpha_1 \right], \nonumber \\
	C_{\zeta^{0,i} H_1^- W^+\gamma} &=& -\frac{i}{2 s_W} \left[ -\frac{n}{2} \sin\alpha_1 + T_{-1}^+ \cos\alpha_1 \right], \nonumber \\
	C_{H_1^Q H_1^{-(Q+1)} W^+\gamma} &=& -\frac{2Q+1}{\sqrt{2}s_W} \left[ T_Q^+ \cos\alpha_Q \cos\alpha_{Q+1} - T_{-Q-1}^+ \sin\alpha_Q \sin\alpha_{Q+1} \right], \nonumber \\
	C_{H_1^Q H_2^{-(Q+1)} W^+\gamma} &=& -\frac{2Q+1}{\sqrt{2}s_W} \left[ -T_Q^+ \cos\alpha_Q \sin\alpha_{Q+1} - T_{-Q-1}^+ \sin\alpha_Q \cos\alpha_{Q+1} \right], \nonumber \\
	C_{H_2^Q H_1^{-(Q+1)} W^+\gamma} &=& -\frac{2Q+1}{\sqrt{2}s_W} \left[ -T_Q^+ \sin\alpha_Q \cos\alpha_{Q+1} - T_{-Q-1}^+ \cos\alpha_Q \sin\alpha_{Q+1} \right], \nonumber \\
	C_{H_2^Q H_2^{-(Q+1)} W^+\gamma} &=& -\frac{2Q+1}{\sqrt{2}s_W} \left[ T_Q^+ \sin\alpha_Q \sin\alpha_{Q+1} - T_{-Q-1}^+ \cos\alpha_Q \cos\alpha_{Q+1} \right], \nonumber \\
	C_{H_1^{n/2-1} \zeta^{-n/2} W^+\gamma} &=& -\frac{n-1}{\sqrt{2}s_W} T^+_{n/2-1} \cos\alpha_{n/2-1}, \nonumber \\
	C_{H_2^{n/2-1} \zeta^{-n/2} W^+\gamma} &=& \frac{n-1}{\sqrt{2}s_W} T^+_{n/2-1} \sin\alpha_{n/2-1}.
\end{eqnarray}
The couplings of two scalars to $W^- \gamma$ are obtained using the relation
\begin{equation}
	C_{s_2^* s_1^* W^- \gamma} = (C_{s_1 s_2 W^+ \gamma})^*.
\end{equation}
Note that all the couplings $C_{s_1 s_2 W^+ \gamma}$ are real except for those that involve one $\zeta^{0,i}$, which are imaginary.

\subsubsection{Couplings to WZ}

The Feynman rule for the coupling $s_1 s_2 W^{\pm}_{\mu} Z_{\nu}$ is given by
\begin{equation}
	-i e^2 C_{s_1 s_2 W^{\pm} Z} g_{\mu\nu},
\end{equation}
where 
\begin{eqnarray}
	C_{\zeta^{0,r} H_1^- W^+Z} &=& -\frac{1}{2 s_W^2 c_W} \left[ -\frac{n}{2} s_W^2 \cos\alpha_1 - (2 - s_W^2) T_{-1}^+ \sin\alpha_1 \right], \nonumber \\
	C_{\zeta^{0,r} H_2^- W^+Z} &=& -\frac{1}{2 s_W^2 c_W} \left[ \frac{n}{2} s_W^2 \sin\alpha_1 - (2 - s_W^2) T_{-1}^+  \cos\alpha_1 \right], \nonumber \\
	C_{\zeta^{0,i} H_1^- W^+Z} &=& -\frac{i}{2 s_W^2 c_W}  \left[ -\frac{n}{2} s_W^2 \cos\alpha_1 + (2 - s_W^2) T_{-1}^+  \sin\alpha_1 \right], \nonumber \\
	C_{\zeta^{0,i} H_2^- W^+Z} &=& -\frac{i}{2 s_W^2 c_W} \left[ \frac{n}{2} s_W^2 \sin\alpha_1 + (2 - s_W^2) T_{-1}^+ \cos\alpha_1 \right], \nonumber \\
	C_{H_1^Q H_1^{-(Q+1)} W^+Z} &=& -\frac{1}{\sqrt{2} s_W^2 c_W}\left[ - (2(Q+1) + (2Q+1) s_W^2) T^+_{-Q-1} \sin\alpha_Q \sin\alpha_{Q+1} \right. \nonumber \\ 
	&& \qquad \qquad \qquad \left. +(2Q c_W^2 - s_W^2) T^+_Q \cos\alpha_Q \cos\alpha_{Q+1} \right], \nonumber \\
	C_{H_1^Q H_2^{-(Q+1)} W^+Z} &=& -\frac{1}{\sqrt{2} s_W^2 c_W}\left[ - (2(Q+1) + (2Q+1) s_W^2) T^+_{-Q-1} \sin\alpha_Q \cos\alpha_{Q+1} \right. \nonumber \\ 
	&& \qquad \qquad \qquad \left. -(2Q c_W^2 - s_W^2) T^+_Q \cos\alpha_Q \sin\alpha_{Q+1}\right], \nonumber \\
	C_{H_2^Q H_1^{-(Q+1)} W^+Z} &=& -\frac{1}{\sqrt{2} s_W^2 c_W}\left[  -(2(Q+1) + (2Q+1) s_W^2) T^+_{-Q-1} \cos\alpha_Q \sin\alpha_{Q+1} \right. \nonumber \\ 
	&& \qquad \qquad \qquad \left. -(2Q c_W^2 - s_W^2) T^+_Q \sin\alpha_Q \cos\alpha_{Q+1}\right], \nonumber \\
	C_{H_2^Q H_2^{-(Q+1)} W^+Z} &=& -\frac{1}{\sqrt{2} s_W^2 c_W}\left[ - (2(Q+1) + (2Q+1) s_W^2) T^+_{-Q-1}\cos\alpha_Q \cos\alpha_{Q+1} \right. \nonumber \\ 
	&& \qquad \qquad \qquad \left. +(2Q c_W^2 - s_W^2) T^+_Q \sin\alpha_Q \sin\alpha_{Q+1}  \right], \nonumber \\
	C_{H_1^{n/2-1} \zeta^{-n/2} W^+Z} &=& -\frac{1}{\sqrt{2} s_W^2 c_W}(n c_W^2-2 + s_W^2) T_{n/2-1}^+ \cos\alpha_{n/2-1}, \nonumber \\
	C_{H_2^{n/2-1} \zeta^{-n/2} W^+Z} &=& \frac{1}{\sqrt{2} s_W^2 c_W} (n c_W^2-2 + s_W^2) T_{n/2-1}^+ \sin\alpha_{n/2-1}.
\end{eqnarray}
The couplings of two scalars to $W^- Z$ are obtained using the relation
\begin{equation}
	C_{s_2^* s_1^* W^- Z} = (C_{s_1 s_2 W^+ Z})^*.
\end{equation}
Note that all the couplings $C_{s_1 s_2 W^+ Z}$ are real except for those that involve one $\zeta^{0,i}$, which are imaginary.



\begin{thebibliography}{99}

\bibitem{SUSY}
P.~Fayet, 
  Phys.\ Lett.\ B {\bf 64}, 159 (1976);
 Phys.\ Lett.\ B {\bf 69}, 489 (1977);
 Phys.\ Lett.\ B {\bf 84}, 421 (1979);
G.~R.~Farrar and P.~Fayet, 
 Phys.\ Lett.\ B {\bf 76}, 575 (1978).

\bibitem{LH}
N.~Arkani-Hamed, A.~G.~Cohen and H.~Georgi,
  Phys.\ Lett.\ B {\bf 513}, 232 (2001)
  [hep-ph/0105239];
N.~Arkani-Hamed, A.~G.~Cohen, E.~Katz, A.~E.~Nelson, T.~Gregoire and J.~G.~Wacker,
  JHEP {\bf 0208}, 021 (2002)
  [hep-ph/0206020];
N.~Arkani-Hamed, A.~G.~Cohen, E.~Katz and A.~E.~Nelson,
  JHEP {\bf 0207}, 034 (2002)
  [hep-ph/0206021];
M.~Schmaltz and D.~Tucker-Smith,
  Ann.\ Rev.\ Nucl.\ Part.\ Sci.\  {\bf 55}, 229 (2005)
  [hep-ph/0502182].

  \bibitem{Cirelli:2005uq} 
  M.~Cirelli, N.~Fornengo and A.~Strumia,
  Nucl.\ Phys.\ B {\bf 753}, 178 (2006)
  [hep-ph/0512090].

\bibitem{Cirelli:2009uv} 
  M.~Cirelli and A.~Strumia,
  New J.\ Phys.\  {\bf 11}, 105005 (2009)
  [arXiv:0903.3381 [hep-ph]].

\bibitem{Cai:2012kt} 
  Y.~Cai, W.~Chao and S.~Yang,
  JHEP {\bf 1212}, 043 (2012)
  [arXiv:1208.3949 [hep-ph]].
  
\bibitem{Babu:2009aq} 
  K.~S.~Babu, S.~Nandi and Z.~Tavartkiladze,
  Phys.\ Rev.\ D {\bf 80}, 071702 (2009)
  [arXiv:0905.2710 [hep-ph]].
  
   \bibitem{Picek:2009is} 
  I.~Picek and B.~Radovcic,
  Phys.\ Lett.\ B {\bf 687}, 338 (2010)
  [arXiv:0911.1374 [hep-ph]];
  K.~Kumericki, I.~Picek and B.~Radovcic,
  Phys.\ Rev.\ D {\bf 84}, 093002 (2011)
  [arXiv:1106.1069 [hep-ph]].

\bibitem{Ren:2011mh} 
  B.~Ren, K.~Tsumura and X.-G.~He,
  Phys.\ Rev.\ D {\bf 84}, 073004 (2011)
  [arXiv:1107.5879 [hep-ph]].
  
\bibitem{McDonald:2013kca} 
  K.~L.~McDonald,
  JHEP {\bf 1307}, 020 (2013)
  [arXiv:1303.4573 [hep-ph]].
  
 \bibitem{Law:2013gma} 
  S.~S.~C.~Law and K.~L.~McDonald,
  Phys.\ Rev.\ D {\bf 87}, 113003 (2013)
  [arXiv:1303.4887 [hep-ph]].
  
\bibitem{Law:2013saa} 
  S.~S.~C.~Law and K.~L.~McDonald,
  JHEP {\bf 1309}, 092 (2013)
  [arXiv:1305.6467 [hep-ph]].

\bibitem{Chen:2012vm} 
  C.-S.~Chen, C.-Q.~Geng, D.~Huang and L.-H.~Tsai,
  Phys.\  Rev.\  D {\bf 87}, 077702 (2013)
  [arXiv:1212.6208 [hep-ph]].
  
\bibitem{McDonald:2013hsa} 
  K.~L.~McDonald,
  JHEP {\bf 1311}, 131 (2013)
  [arXiv:1310.0609 [hep-ph]].
    
\bibitem{AbdusSalam:2013eya} 
  S.~S.~AbdusSalam and T.~A.~Chowdhury,
  JCAP {\bf 1405}, 026 (2014)
  [arXiv:1310.8152 [hep-ph]].
   
\bibitem{Hisano:2013sn} 
  J.~Hisano and K.~Tsumura,
  Phys.\ Rev.\ D {\bf 87}, 053004 (2013)
  [arXiv:1301.6455 [hep-ph]];
  S.~Kanemura, M.~Kikuchi and K.~Yagyu,
  Phys.\ Rev.\ D {\bf 88}, 015020 (2013)
  [arXiv:1301.7303 [hep-ph]].

\bibitem{Hally:2012pu} 
  K.~Hally, H.~E.~Logan and T.~Pilkington,
  Phys.\ Rev.\ D {\bf 85}, 095017 (2012)
  [arXiv:1202.5073 [hep-ph]].
  
\bibitem{ATLAS:charged} 
  G.~Aad {\it et al.}  [ATLAS Collaboration],
  Phys.\ Lett.\ B {\bf 722}, 305 (2013)
  [arXiv:1301.5272 [hep-ex]].

\bibitem{Kumericki:2012bf} 
  K.~Kumericki, I.~Picek and B.~Radovcic,
  JHEP {\bf 1207}, 039 (2012)
  [arXiv:1204.6597 [hep-ph]].
  
\bibitem{Earl:2013jsa} 
  K.~Earl, K.~Hartling, H.~E.~Logan and T.~Pilkington,
  Phys.\  Rev.\  D {\bf 88}, 015002 (2013)
  [arXiv:1303.1244 [hep-ph]].
  
\bibitem{Peskin:1990zt} 
  M.~E.~Peskin and T.~Takeuchi,
  Phys.\ Rev.\ Lett.\  {\bf 65}, 964 (1990);
  Phys.\ Rev.\ D {\bf 46}, 381 (1992).

\bibitem{Lavoura:1993nq} 
  L.~Lavoura and L.-F.~Li,
  Phys.\ Rev.\ D {\bf 49}, 1409 (1994)
  [hep-ph/9309262];
  H.-H.~Zhang, W.-B.~Yan and X.-S.~Li,
  Mod.\ Phys.\ Lett.\ A {\bf 23}, 637 (2008)
  [hep-ph/0612059].
  
\bibitem{Baak:2012kk} 
  M.~Baak, M.~Goebel, J.~Haller, A.~Hoecker, D.~Kennedy, R.~Kogler, K.~Moenig and M.~Schott {\it et al.},
  Eur.\ Phys.\ J.\ C {\bf 72}, 2205 (2012)
  [arXiv:1209.2716 [hep-ph]].
    
\bibitem{HHG}
J.~F.~Gunion, H.~E.~Haber, G.~L.~Kane, and S.~Dawson, 
{\it The Higgs Hunter's Guide} (Westview, Boulder, 2000).

\bibitem{hgaga-ATLAS}
ATLAS Collaboration, ATLAS-CONF-2013-012 (2013), available from
\url{http://cds.cern.ch/record/1523698}.

\bibitem{hgaga-CMS}
CMS Collaboration, CMS-PAS-HIG-13-001 (2013), available from
\url{http://cds.cern.ch/record/1530524}.

\bibitem{hgz-ATLAS}
ATLAS Collaboration, ATLAS-CONF-2013-009 (2013), available from
\url{http://cds.cern.ch/record/1523683}.

\bibitem{hgz-CMS}
  S.~Chatrchyan {\it et al.}  [CMS Collaboration],
  Phys.\ Lett B {\bf 726}, 587 (2013)
  [arXiv:1307.5515 [hep-ex]].

\bibitem{HGZ-Chen} 
  C.-S.~Chen, C.-Q.~Geng, D.~Huang and L.-H.~Tsai,
  Phys.\ Rev.\ D {\bf 87}, 075019 (2013)
  [arXiv:1301.4694 [hep-ph]].
	
\bibitem{HGZ-Carena} 
  M.~Carena, I.~Low and C.~E.~M.~Wagner,
  JHEP {\bf 1208}, 060 (2012)
  [arXiv:1206.1082 [hep-ph]].
	
\bibitem{Steigman:2012nb} 
  G.~Steigman, B.~Dasgupta and J.~F.~Beacom,
  Phys.\ Rev.\ D {\bf 86}, 023506 (2012)
  [arXiv:1204.3622 [hep-ph]].
  
 \bibitem{Belyaev:2012qa} 
  A.~Belyaev, N.~D.~Christensen and A.~Pukhov,
  Comput.\ Phys.\ Commun.\  {\bf 184}, 1729 (2013)
  [arXiv:1207.6082 [hep-ph]].

\bibitem{Ellis:2000ds} 
  J.~R.~Ellis, A.~Ferstl and K.~A.~Olive,
  Phys.\ Lett.\ B {\bf 481}, 304 (2000)
  [hep-ph/0001005].

\bibitem{Hisano:2011cs} 
  J.~Hisano, K.~Ishiwata, N.~Nagata and T.~Takesako,
  JHEP {\bf 1107}, 005 (2011)
  [arXiv:1104.0228 [hep-ph]].

\bibitem{Akerib:2013tjd} 
  D.~S.~Akerib {\it et al.}  [LUX Collaboration],
  Phys.\ Rev.\ Lett.\  {\bf 112}, 091303 (2014)
  [arXiv:1310.8214 [astro-ph.CO]].
  
\bibitem{Akerib:2012ys} 
  D.~S.~Akerib {\it et al.}  [LUX Collaboration],
  Nucl.\ Instrum.\ Meth.\ A {\bf 704}, 111 (2013)
  [arXiv:1211.3788 [physics.ins-det]].
  
\bibitem{Boulay:2012hq} 
  M.~G.~Boulay [DEAP Collaboration],
  J.\ Phys.\ Conf.\ Ser.\  {\bf 375}, 012027 (2012)
  [arXiv:1203.0604 [astro-ph.IM]].
  
\bibitem{Aprile:2012zx} 
  E.~Aprile [XENON1T Collaboration],
  arXiv:1206.6288 [astro-ph.IM].

\bibitem{DMTools}
DMTools website, \url{http://dmtools.brown.edu}.

\bibitem{Cushman:2013zza} 
  P.~Cushman {\it et al.},
  arXiv:1310.8327 [hep-ex].
  


\end{thebibliography}
\end{document}